# Current challenges for preseismic electromagnetic emissions: shedding light from micro-scale plastic flow, granular packings, phase transitions and self-affinity notion of fracture process.

## K. Eftaxias[1] and S. M. Potirakis[2]

[1] Department of Physics, Section of Solid State Physics, University of Athens, Panepistimiopolis, GR-15784, Zografos, Athens, Greece, ceftax@phys.uoa.gr

[2] Department of Electronics Engineering, Technological Education Institute (TEI) of Piraeus, 250 Thivon & P. Ralli, GR-12244, Aigaleo, Athens, Greece, spoti@teipir.gr .

**Abstract**

Are there credible electromagnetic (EM) EQ precursors? This a question debated in the scientific community and there may be legitimate reasons for the critical views. The negative view concerning the existence of EM precursors is enhanced by features that accompany their observation which are considered as paradox ones, namely, these signals: (i) are not observed at the time of EQs occurrence and during the aftershock period, (ii) are not accompanied by large precursory strain changes, (iii) are not accompanied by simultaneous geodetic or seismological precursors and (v) their traceability is considered problematic. In this work, the detected candidate EM precursors are studied through a shift in thinking towards the basic science findings relative to granular packings, micron-scale plastic flow, interface depinning, fracture size effects, concepts drawn from phase transitions, self-affine notion of fracture and faulting process, universal features of fracture surfaces, recent high quality laboratory studies, theoretical models and numerical simulations. Strict criteria are established for the definition of an emerged EM anomaly as a preseismic one, while, precursory EM features, which have been considered as paradoxes, are explained. A three-stage model for EQ generation by means of preseismic fracture-induced EM emissions is proposed. The claim that the observed EM precursors may permit a real-time and step-by-step monitoring of the EQ generation is tested.

## 1. Introduction

Earthquakes (EQs) possess strong relevance to material science. In a simplified view, an EQ may be regarded as the rubbing of a fault. The way in which a frictional interface fails is crucial to our fundamental understanding of failure processes in fields ranging from engineering to the study of EQs, e.g., (Kawamura, 2012, and references therein). Understanding how EQs occur is one of the most challenging questions in fault and EQ mechanics (Shimamoto and Togo, 2012).

The main effort has been devoted to the study of EQ dynamics at the laboratory scale. Two different directions have been followed for this purpose. The first one mainly focuses on the understanding of the laws that govern friction (e.g., Johnson et al., 2008; Zapperi, 2010; Ben-David et al., 2010; Chang et al., 2012). The second one refers to the fracture induced acoustic emission (AE) and electromagnetic emission (EME) techniques, as they are sensitive to the micro-structural changes occurring in the sample. Indeed, it has been found that micro- and macro-cracking processes are accompanied with EME and AE ranging in a wide frequency spectrum, i.e., from the kHz band to the MHz band. Especially, insight into EQ dynamics has been gained through fracture studies of usually pre-cut rock samples, to mimic slip on pre-existing tectonic faults (Lockner et al., 1991). Such studies have established that there is a considerable overlap between the statistics of EQs and laboratory AE / EME studies. These include basic properties such as the Gutenberg-Richter frequency-magnitude relation and the correspondence of Omori's law for aftershocks and primary creep decay sequences (Lockner, 1996; Baddari et al., 1999; Rumi and Ananthakrishna, 2004 and references therein; Fukui et al., 2005; Kumar and Misra, 2007; Chauhan and Misra, 2008; Baddari and Frolov, 2010; Baddari et al., 2011; Lacidogna et al., 2010; Schiavi et al., 2011; Carpinteri et al., 2012).

It is practically impossible to install an experimental network to measure stress and strain at the location where an EQ is generated (focus area) using the same instrumentation as in laboratory experiments. It is therefore impossible *to investigate the corresponding states of stress and strain and their time variation* in order to understand the laws that govern the last stages of EQ generation, or to monitor (much less to control) the principal characteristics of a fracture process. In principle, this disadvantage does not accompany the tool of the fracture induced EME in the case of significant surface EQs that occur on land, keeping in mind that laboratory experiments are man controlled while field observations are measurements of events over which researchers have no control. On the contrary, the EME method is expected to reveal more information when it is used at the geophysical scale. Indeed, a major difference between the laboratory and natural processes is the order-of-magnitude differences in scale (in space and time), allowing the possibility of experimental observation at the geophysical scale for a range of physical processes which are not observable at the laboratory scale (Main and Naylor, 2012). At the laboratory scale the fault growth process normally occurs violently in a fraction of a second (Lockner et al., 1991). Thus, the idea that field observations at the geophysical scale by means of EME will probably reveal features of the last stages of failure process which are not clearly observable at the laboratory scale, allowing the monitoring in real-time and step-by-step of the gradual damage of stressed materials during EQ preparation process, cannot, in principle, be excluded.

Based on the above mentioned expectation we installed a field measurement network using the same instrumentation as in laboratory experiments for the recording of fracture-induced kHz and MHz magnetic and electric fields, respectively. Since 1994, a telemetric remote station has been installed in a carefully selected mountainous site of Zante island at the south-west of the island (37.76$^\circ$ N–20.76$^\circ$ E) providing low EM background noise. The complete measurement system comprises of (i) six loop antennas detecting the three components (EW, NS, and vertical) of the variations of the magnetic field at 3 kHz and 10 kHz respectively; (ii) three vertical $\lambda/2$ electric dipole antennas detecting the electric field variations at 41, 54 and 135 MHz respectively; (iii) other magnetic and electromagnetic sensors. All the time-series are sampled once per second, i.e., with a sampling frequency of 1 Hz. The block diagram of the measurement configuration is shown in Fig. 1. Note that the main focus is on the recorded MHz and kHz EME. The measured frequencies (3 kHz, 10 kHz, 41 MHz, 54 MHz and 135 MHz) were selected in order to minimize the effects of the man-made noise in the mountainous area of Zante. *We note that the installed experimental setup helps us not only to specify whether or not a single MHz or kHz EM anomaly is preseismic in itself, but also whether a sequence of MHz and kHz EM disturbances which emerge one after the other in a short time period, could be characterized as preseismic one.*

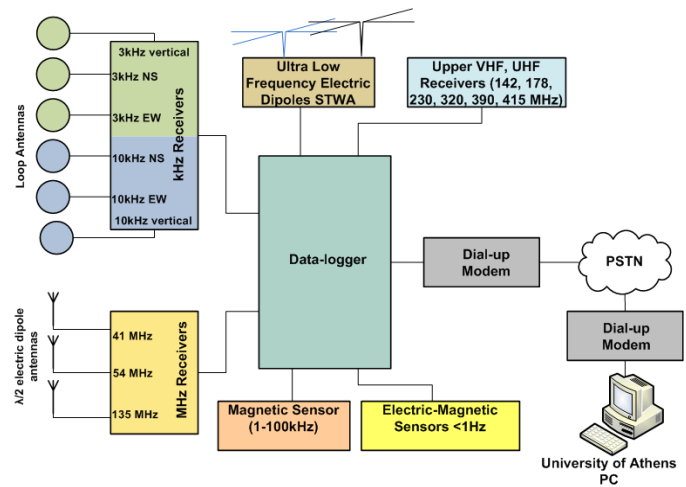

**Fig. 1** The block diagram of the measurement configuration installed at the Zante remote telemetric station. The recorded data are acquired and regularly forwarded to the central telemetric station of the University of Athens through PSTN.

*Are there credible EM earthquake precursors?* This is a question debated in the scientific community (Uyeda, 2009). There may be legitimate reasons for the critical views. The degree to which we can establish an EME as a precursory phenomenon reflecting the underlying failure process is depended on how well we understand the failure processes. However, many aspects of EQ generation still escape our full understanding. "No scientific prediction is possible without exact definition of the anticipated phenomenon and the rules, which define clearly in advance of it whether the prediction is confirmed or not" (Kossobokov, 2006). It is impossible to identify a recorded EM anomaly as preseismic one by inspection! Indeed, "there was a lot of bad science calling itself prediction" (Cyranoski, 2004). Thus, earthquake researchers armed with arguments to invalidate the word "prediction" appeared (Wyss, 1997; Geller et al., 1997a).

The negative view concerning the existence of EM precursors is enhanced by paradox features that accompany their observation. More precisely:

1. A common observation from all of the experiments designed to detect fracture induced EM precursors at the geophysical scale is that MHz-kHz signals are observed before EQs, but these signals are not observe at the time of EQs occurrence, namely, *an EM silence systematically emerges before the time of the EQ occurrence* (Gokhberg et al., 1995; Matsumoto et al., 1998; Hayakawa Fujinawa, 1994; Hayakawa, 1999; Morgounov, 2001; Eftaxias et al., 2002 and references therein; Eftaxias et al., 2012) . The following relevant argument is often raised by some scientists leading them to the conclusion that the observed kHz EM anomalies are not preseismic ones, despite the existence of any other strong multidisciplinary documentation supporting their validity (e.g., Park et al., 1993; Geller, 1997; Geller et al., 1997b; Johnston, 1997): If the recorded EM signals are emitted from micro-fracturing, why isn't any EM signal detected at the time of the EQ occurrence? We note that laboratory experiments *by means of AE* detect the largest emission at the collapse of rock samples, while the AE and EME were for a long time considered as two faces of the same coin. Thus, really, it seems that there is no reason why there is a time gap between the observed EM anomalies and the occurrence of the EQ. Note, however, that newer evidence shows that the view that AE and EME are two faces of the same coin is not generally valid. It is valid only during the plastic flow (damage) phase, but not during the last phase of the post-peak stage (see Sections and 5.1 and 6).

2. EM emissions are not observed during the aftershock period. This feature is considered as enigmatic one, as well.

3. Strain changes are largest at the time of EQ. The general observation of EM precursory signals without co-seismic ones is considered a paradox on the grounds that any mechanism must explain why the emerged EM signals are not accompanied by large precursory strain changes, much larger from co-seismic ones. Geller et al. (1997a) emphasize that the absence of simultaneous *geodetic or seismological* precursors means that the observed EM anomalies are not preseismic ones. In fact, a candidate precursory EM activity should be consistent with other precursors that are imposed by data from other disciplines.

4. The traceability of EM precursors is also considered as problematic on the grounds that they should normally be absorbed by the Earth's crust.

*The scope of the present work.* The above mentioned legitimate critic views suggest that the observed EM precursors should be formulated through a shift in thinking towards the basic science findings of fracture / faulting process. We attempt to formulate these precursors based on recent high quality laboratory studies, theoretical models and numerical simulations relative to granular packings, micron-scale plastic flow, fracture size effects, interface depinning, self-affine notion of fracture and faulting process, concepts drawn from phase transitions, and universal features of fracture surfaces. Especially, we think that it is useful to consider the sublevel (or microscopic) ingredients that underlie the macroscopic phenomenology of failure process in order to understand the former paradoxes (Papanikolaou et al., 2012). Recent progress in experimental techniques, allowing one to test and probe materials at sufficiently small length, or time scales, or in three dimensions, has led to a quantitative understanding of the physical processes involved from the micro- to the geophysical scales (Bouchaud and Soukiassian, 2009; Papanikolaou et al., 2012). In parallel, simulations have also progressed; by extending the time and length scales they are able to reach and thus to attain conditions experimentally accessible at the geophysical scale. Thus, one can study the candidate EM precursors under the light of these new progresses. Such a study had not been previously attempted. We pay attention to recent studies (Papanikolaou et al., 2012 and references therein) revealing that "single microcrystals display a rich collection of novel mechanical behaviors: together with size effects and the emergence of avalanche slip events, the importance of often-neglected slow processes on intermittency has now come to light. The presented experiments at the microscale now force us to reconsider our understanding of the macroscopic world, such as disordered solids and earthquake faults."

Based on the above mentioned concepts, we try to establish strict criteria for the definition of an emerged EM anomaly as a preseismic one through a proposed *three stages model of EQ generation* by means of the observed EM precursors. In the frame of the above mentioned direction, our effort has been recently focusing, in an appropriately critical spirit, on asking five basic questions: *(i) How can we construct a strict set of criteria which will permit to discriminate a sequence of emerged MHz and kHz EME as a preseismic one? (ii) How can we link an individual MHz and kHz EM precursor with a distinctive stage of the EQ preparation? This question is the most crucial one. (iii) Is the systematically observed (for a long time and globally) EM silence before the EQ occurrence a paradox feature or does it constitutes the last precursor proclaiming that the under preparation EQ is imminent? (iv) How can we identify key symptoms in an EM precursor which signify that the occurrence of the prepared EQ is unavoidable? (v) Are the considered as enigmatic features enigmatic indeed?* Finally, the claim that the observed EM precursors may permit a real-time and step-by-step monitoring of the EQ generation is also tested.

## 2. A three-stage model for earthquake generation by means of preseismic fracture-induced EM emissions.

Two important features are consistently observed at the geophysical scale:

(i) The launch of the preseismic MHz radiation systematically precedes the kHz one (Gokhberg et al., 1995; Matsumoto et al., 1998; Hayakawa Fujinawa, 1994; Hayakawa, 1999; Morgounov, 2001; Kapiris et al., 2004; Eftaxias et al., 2002, and references therein). This situation is also observed at the laboratory scale by means of AE and EME. At the geophysical scale the MHz EME often emerges during the last week before the EQ while the kHz EME from half an hour up to a few decades of hours before the shock. (ii) As it has been mentioned, EM signals are not observed at the time of EQ occurrence. Based on the above mentioned experimental features we propose the following three stage model for EQ generation by means of preseismic fracture-induced MHz-kHz EMEs.

**Stage 1.** A significant EQ is what happens when the two surfaces of a major fault slip one over the other under the stresses rooted in the motion of tectonic plates. However, large stresses siege the major fault after the activation of a population of smaller faults in the heterogeneous region that surrounds the major fault (Fig. 2). Seismicity triggering is driven by the smallest EQs which trigger fewer events than larger EQs, however they are much more numerous (Helmstetter, 2003).

It has been proposed that the initially observed preseismic MHz EM emission (Fig. 3) is due to the fracture of the highly heterogeneous system that surrounds the formation of strong brittle and high-strength entities (asperities) distributed along the rough fault surfaces sustaining the system. This emission shows antipersistent behavior and can be described in analogy with a phase transition of second order in equilibrium (Kapiris et al., 2004; Contoyiannis et al., 2005; Contoyiannis and Eftaxias, 2008; Contoyiannis et al., 2010; Eftaxias et al., 2007, 2012; Potirakis et al., 2013).

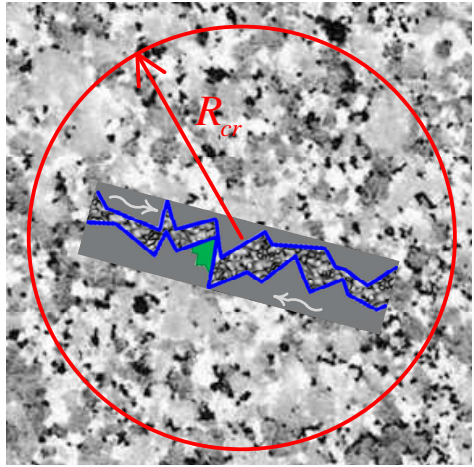

**Fig. 2** A fault (blue lines) is embedded in an heterogeneous environment. The EQ preparation process at the first stage concerns the fracture of a disorder medium surrounding over a critical circle (Bowman et al., 1998; Sammis and Sornette, 2002) the major fault emitting the MHz EME which can be described by means of a phase transition of second order. The symmetry breaking signalizes the transition from the phase of non-directional, almost symmetrical, cracking distribution to a directional localized cracking zone along the direction of the fault. The EQ is inevitable if and only if the asperities break (green highlighted area), emitting the kHz EME during the second stage, and then an EME silence follows.

**Stage 2**. Laboratory experiments of rock fracture and frictional sliding have shown that the relative slip of two fault surfaces takes place in two phases. *A stick-slip like fracture-sliding precedes dynamical fast global slip* (Bouchon et al., 2001; Baumberger et al., 2002; Rubinstein et al., 2004, 2007; Chang et al., 2012; Kammer et al. 2012). Our understanding of different rupture modes is still very much in its infancy (Ben-David et al., 2010). We note that the pioneering laboratory friction experiments of Rabinowicz (1951) showed that the transition between static and dynamic friction occurs over a characteristic slip. Recent studies reveal that physical systems under slowly increasing stress may respond through abrupt events. Such jumps in observable quantities are abundant, being found in systems ranging from complex social networks to EQs (Papanikolaou et al., 2012). We propose that the abruptly emerging sequence of kHz EM avalanches (Fig. 4 and Fig. 5) originates in the stage of stick-slip-like plastic flow. This precursor shows persistent behavior and does not include any signature of phase transition of second order in equilibrium.

**Stage 3.** We propose that the systematically observed EM silence in all frequency bands is sourced in the stage of preparation of dynamical slip which results to the fast, even super-shear, mode that surpasses the shear wave speed (Ben-David et al., 2010).

In the following sections we present arguments that support the aforementioned hypothesis.

## 3. Stage 1: Does the emergence of the MHz EME reveal the fracture of the highly heterogeneous system surrounding the major fault? - An interpretation in terms of criticality.

The compressive failure of a disordered medium appears as a complex cumulative process involving long-range correlations, interactions, and coalescence of microcracks. Nature seems to paint the following picture in the fracture of a heterogeneous system (Garcimartin et al., 1997; Bowman, 1998; Sornette, 2000; Girard et al., 2010; 2012; Halasz et al., 2012). In the early stages of deformation, when the disorder medium is subjected to external load, the weak components break immediately and serve as nucleation centers for the growth of broken clusters. The load transferred to the nearest neighbors of broken components gives rise to further breaking. As deformation proceeds cooperative effects appear, cracking areas cluster in space according to scale-free patterns and are dynamically interacting to each other. As the external load increases larger clusters are formed and long-range correlations buildup through local interactions until they extend throughout the entire system. At peak load, the largest damage cluster does not yet span the heterogeneous system (Girard et al., 2012). Then, during the post-peak phase cracking events are localized in the vicinity of one or a few large damage clusters that eventually evolve into a spanning cluster. Girard et al. (2010) using simulations conclude that the spatial correlation length associated with damage events reaches the size of the system at peak load. *This means that the divergence of the correlation length precedes the final failure of the disorder medium.* Strain-driven, compressive failure experiments on rocks have resulted in a similar observation: the failure plane is not fully formed at peak load (Lockner et al., 1991). *All these results advocate for a critical point interpretation of failure.* The challenge in the analysis of a recorded MHz EM time-series is to show that this includes the above mentioned features and especially to detect the "critical epoch" during which the "short-range" correlations evolve into "long-range" ones, as well as the epoch of localization of damage. We argue that the aforementioned two crucial epochs can be identified in the recorded MHz EM time series. More precisely:

Based on a fractal spectral analysis of the MHz EM time series, it has been shown that the associated Hurst-exponent, $H$, lies in the range $0 < H < 0.5$ indicating that the dynamics of the observed MHz EM field is characterized by *anti-persistency* (Kapiris et al., 2004; Contoyiannis et al., 2005), namely, if the EM fluctuations increase in one period, it is likely to continuous in the period immediately following, and vice-versa. This means that the underlying fracture mechanism is characterized by a negative feedback mechanism that "kicks" the cracking rate away from extremes. The existence of antipersistency supports the suspicion that *the MHz EME would be described in analogy with a thermal second order phase transition in equilibrium* (Contoyiannis et al., 2005). This really happens. Indeed:

Characteristic features at a *critical point* of a second order transition are: (i) the existence of strongly correlated fluctuations, right at the "critical point" the subunits / cracking clusters are well correlated even at arbitrarily large separation, namely, the correlation function $C(r)$ follows long-range power-law decay; (ii) the appearance of self-similar structures both in time and space. This fact is mathematically expressed through power law expressions for the distributions of spatial or temporal quantities associated with the aforementioned self-similar structures (Stanley, 1987, 1999). *Below and above of the critical point a dramatic breakdown of critical characteristics, in particular long-range correlations, appears; the correlation function turns into a rapid exponential decay (Stanley, 1987, 1999).*

Recently the method of critical fluctuations (MCF) has been introduced, which can reveal the critical state as well as the departure from critical state (Contoyiannis and Diakonos, 2000; Contoyiannis et al., 2002). The analysis of MHz EME by means of the MCF reveals (see Fig. 3):

> (i) The time-window in the EM time-series that corresponds to the *"critical window"*, namely, the epoch during which the short-range correlations between the cracking areas have been evolved to long-range ones (Contoyiannis et al., 2005; 2010; 2013). More precisely, *the laminar lengths (waiting times) fit a power-law type distribution with exponents usually approaching the value 1.4*. Importantly, the *"critical window" in the MHz time series is characterized by strong antipersistency* (Contoyiannis et al., 2005; 2010; 2013). *The associated physical information is that the control mechanism regulating the fracture kicks the system out of extreme situations providing adaptability, the ability to respond to various external stresses.*

We note that, Halasz et al. (2012), based on a fiber bundle model of subcritical fracture with localized load sharing, showed that for high disorder, simultaneously growing cracks are spread homogeneously over the entire disordered system, while the distribution of waiting times follows a power-law functional form with exponent 1.4. As it is said, the spatial correlation length associated with damage clusters reaches the size of the system at peak load (Girard et al., 2010; 2012).

> (ii) The "*non critical window*" in the MHz EM time series which is emerged after the appearance of critical window (Contoyiannis et al., 2005; 2010; 2013). The timescale invariance that characterizes the critical window has been lost; *the laminar lengths (waiting times) fit an exponential type distribution.* This means that short-range correlations between the cracking areas have been emerged. Moreover, this window shows lower antipersistency, namely it becomes less anti-correlated, as the associated $H-$exponents are closer to 0.5; the system has lost a part of its adaptability, namely, the ability to respond to stresses.

The above mentioned transition from the critical epoch to the non critical one constitutes a crucial feature of second order phase transition known as *"symmetry breaking"* (Contoyiannis et al., 2005): its appearance reveals the transition from the phase of non-directional, almost symmetrical, cracking distribution to a directional localized cracking zone. Therefore, the MCF also reveals the time window of the MHz EM time-series where the emissions of the post-peak phase are strongly localized along the main fault (Lockner et al., 1991; Girard et al., 2010; 2012). The integration of the "symmetry breaking" implies that the rupture process has already been obstructed along the backbone of strong asperities sustaining the fault surfaces. The "siege" of asperities has already been started. *However, this does not mean that the EQ is unavoidable. The abrupt emergence of strong avalanche-like kHz EM activity reveals the fracture of asperities, if and when the local stresses exceed their fracture stress* (Contoyiannis et al., 2005).

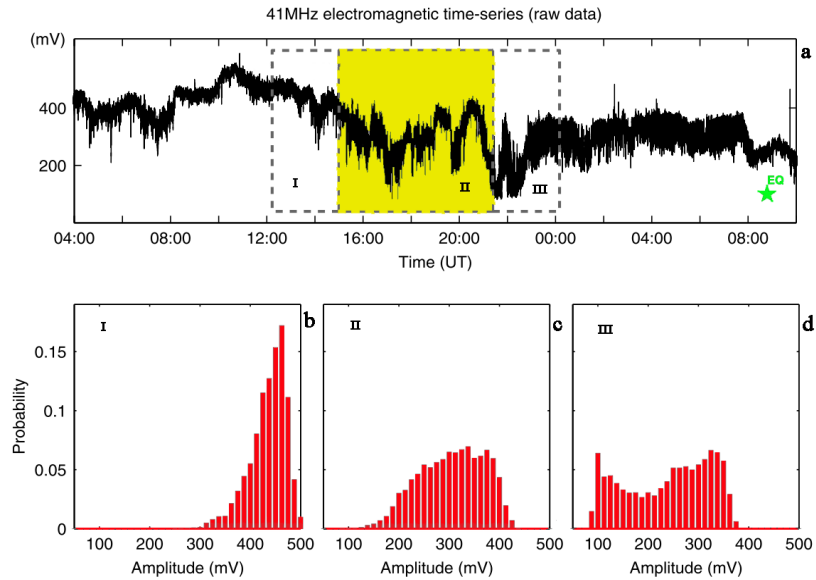

**Fig. 3** (a) The 41-MHz time-series associated with the Kozani-Grevena EQ ($M_W = 6.6$, 13 May 1995). The green star indicates the time of the EQ occurrence. (b) - (d) show the distribution of the amplitude of electromagnetic pulses for three consecutive time intervals marked in (a). The second (yellow shaded) time interval determines, in terms of the method of critical fluctuations (MCF), the crucial time interval during which the short-range correlations evolve to long range (critical window); the corresponding distribution (c) might be considered to be a precursor of the impending symmetry breaking readily observable in the subsequent time interval (d). The aforementioned evolution is expected in the framework of the hypothesis that the fracture in the highly disordered media develops as a kind of generalized continuous phase transition.

A crucial question refers to what is the physical mechanism that organizes the heterogeneous system in its critical state. Lévy fights and Lévy walks are applied in modeling physical systems with spatiotemporal fractality (Bouchaud and Georges, 1990). The characteristic feature of Lévy fight is that it does not converge to the Gaussian stochastic process; instead it is "attracted" towards the Lévy stable process with infinite variance. Lévy stable distributions, although they play an important role in mathematics are basically non-physical, because in real world there exist no processes ("Lévy flights") which would produce empirical data with infinite moments (Kwapień and Drożdża, 2012). For this reason, a family of more realistic distributions called *"truncated Lévy distributions"* was introduced by Mantegna and Stanley (1994) where an upper cutoff to the values of random variables was introduced. Combining ideas of: *truncated Lévy* statistics, nonextensive Tsallis statistical mechanics, and criticality with features hidden in the precursory MHz time-series we have shown that a truncated Lévy walk type mechanism can organize the heterogeneous system to criticality (Contoyiannis and Eftaxias, 2008 and references therein). Intuitively, the proposed Lévy walk mechanism could be the result of a feedback "dialogue" between the stresses and heterogeneity (Contoyiannis and Eftaxias, 2008).

In summary, it might be difficult for someone to accept that such MHz EM anomalies are indeed seismogenic ones. However it is even more difficult to prove that they are not. The hypothesis that the analysis of MHz EM time-series permits the step-by-step monitoring of the time evolution of the fracture of disordered material surrounding the major fault in the stressed region cannot be excluded. The emergence of MHz EM silence before the time of the main seismic shock occurrence is justified, as well.

## 4. Stage 2: Does the emergence of the strong avalanche-like kHz EME reveal the sticking frictional stage?

The answer to the question under study, due to its crucial character, requires a thorough documentation. A set of strong criteria should be formed which will permit a strict test of the hypothesis that the kHz EME originates in the damage of asperities. In this direction we refer to: (i) recent laboratory experiments; (ii) basic aspects of faulting and fracture widely documented, especially focusing on the self-affine nature and universal characteristics of these processes; and (iii) features which characterize an extreme event.

### 4.1 Focus on laboratory experiments.

A magnified view of fault surfaces reveals a rough looking surface with high asperities and low valleys. Thus, the macroscopic frictional behavior is mainly determined by the rheological properties of asperities; two surfaces in sliding motion will contact first at these high asperities (Åström et al., 2000). Because of the stress concentration at asperities, molecules or atoms are directly pushed into contact so that an asperity may be viewed as a grain boundary, possibly with some inclusions and impurities (Kawamura et al, 2012 and references therein).

However, how is the family of asperities broken? The following model is widely accepted (Chang et al., 2012 and references therein). Large EQs initiate at a small nucleation area and grow as propagating rupture fronts (Reches, 1999). The propagating fronts activate a multitude of fault patches (asperities) that undergo intense deformation [see Fig. 1 in (Chang et al., 2012)]. Before the front arrives, the stress on each patch is generally lower than its static strength. If the arriving front raises the stress to the static strength level, the patch strength may drop and it slips releasing elastic energy stored in the rocks, and eventually it decelerates and stops. In this way, the frictional fault surfaces suddenly slip, lock and then slip again in a repetitive manner forming the *"stick-slip" state*. The discovery of stick-slip phenomena has revolutionized our understanding of how faults accommodate relative plate motions (Peng and Gomberg, 2010).

It is reasonable to accept that during the damage of a patch a part of the released energy is emitted in the form of an EM avalanche (an "*electromagnetic earthquake*", EM-EQ). We note that the greater the index of brittleness, compressive strength, elastic moduli, and volume of the damaged asperity, the greater is the EME energy generated (Wang and Zhao, 2013).

*In the frame of this hypothesis, the sequence of "electromagnetic EQs" included in an abruptly emerged avalanche-like strong kHz EM emission associated with the activation of a single fault mirrors the sequential damage of asperities / sequence of "stick-slip" events that characterize the stage of quasi-static stick-slip-like sliding (see Fig. 4 and Fig. 5).*

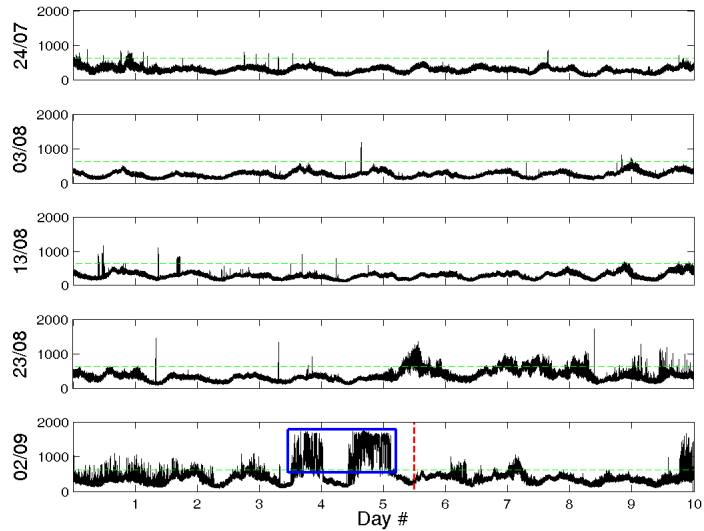

**Fig. 4** Recordings of the magnetic field strength (in arbitrary units) from Zante station at 10 kHz (direction East-West) for the time period from 24 July 1999 to 12 September 1999. The vertical red dashed line denotes the exact time of the Athens' EQ occurrence ($M_W = 5.9$, 7 Sept. 1999). The horizontal green dashed line indicates the considered background noise level. The blue frame (see Fig. 5a for a zoom on this part of the recordings) shows the couple of persistent strong burst-like EM emissions emitted just before the EQ occurrence. Note that fault model of this EQ predicts the activation of two faults with energy relation 80% to 20% (Kontoes et al., 2000). Interestingly, the second EM anomaly, contains approximately 80% of the total EM energy released, with the first one containing the rest 20%, while the same distribution holds for the entropy and information content of these two strong EM anomalies (see also Fig. 6) (Potirakis et al., 2012b).

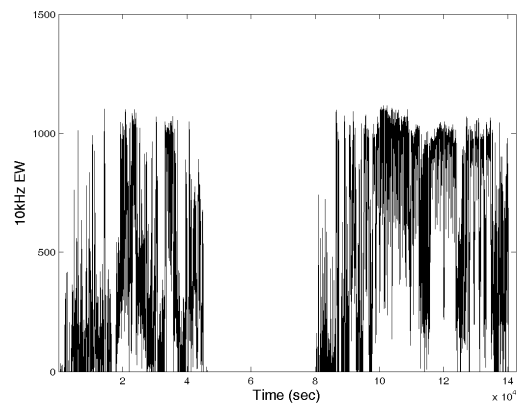

(a)

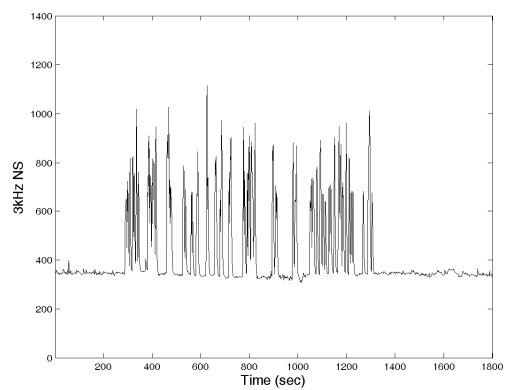

(b)

**Fig. 5** Recordings of the magnetic field strength (in arbitrary units) from Zante station: strong burst-like EM emissions emitted just before the EQ occurrence (a) at 10kHz (direction East-West) before the Athens' EQ ($M_W = 5.9$, 7 Sept. 1999) (above the background noise level, please refer to the blue frame in Fig. 4), and (b) at 3kHz (direction North-South) before the Kozani-Grevena EQ ($M_W = 6.6$, 13 May 1995).

Recent high quality laboratory experiments strongly support the above mentioned crucial hypothesis. Indeed, stick-slip events rooted in the damage of a strong contacts are characterized by sudden shear stress drops that range from 10%-30% of the maximum frictional strength (Johnson et al., 2008). Thus, if our hypothesis is correct, laboratory experiments should reveal that the radiation of EM signals is observed only during sharp drops in stress close to global rupture. This situation really happens. *Strong EM pulses are emitted only during stress drops around the peak of stress, while the amplitude of the emitted EM fields is proportional to the rate of stress drop. On the contrary, during the last phase of the post-peak stage, that is the softening branch in the load versus time diagram, no EME is detected, while, in contrast, the most intense AE emerges* (Fukui et al., 2005; Lacidogna et al., 2011; Carpinteri et al., 2011; Carpinteri et al., 2012). *Consequently, a kHz EM silence is observed at the laboratory scale as it is observed at the geophysical scale.* Importantly, Tsutsumi and Shirai (2008) recorded clear transients in the electric and magnetic fields upon sudden slip in stick-slip experiments on dry quartz-free rock specimen. The authors note that the transients EM signals were observed only when the fault slip occurred by stick-slip mode, not by a stable mode of the sliding, and the amplitudes of signals increased with increasing stress drop. The authors propose that the generation process of the EM signals is closely related to the characteristic behavior of the fault at the time of the initiation of slip during stick-slip events, probably with respect to the intensity of the signals.

A characteristic feature of the observed kHz EM precursor is its *abrupt launch*. This feature is well justified in terms of laboratory and numerical experiments. Indeed, as it was mentioned, large EQs initiate at a small nucleation area and grow as propagating rupture fronts which activate the multitude of fault patches. Numerical (e.g. Lockner and Madden, 1991) and laboratory (e.g. Reches, 1999; Reches and Lockner, 1994) studies indicate the *abrupt initiation of the nucleation phase* of the ensuing global rupture justifying the abrupt emergence of kHz EM precursor. The abrupt cease of the kHz EME is also justified (see Section 5.1).

We note that, McGarr and Fletcher (2003) and McGarr et al. (2010) suggest that stick-slip friction events observed in the laboratory and EQs in continental settings, even with large magnitudes, have similar rupture mechanisms.

**4.2 Focus on the aspect of the self-affine feature of fracture and faulting.**

From the early work of Mandelbrot, the aspect of the self-affine nature of faulting and fracture is widely documented from field observations, laboratory experiments, and studies of failure precursors at the small (laboratory) and large (geophysical) scale. Universal structural patterns of fracture surfaces, weakly dependent on the nature of the material, on the failure mode, and on the scale of fracture, have been well established. *Therefore, a set of strong criteria has been formed that permit a strict test of the hypothesis that the kHz EME originates in the damage of asperities.* We have shown that such a test is positive. Indeed:

Fracture surfaces were found to be self-affine following the persistent fractional Brownian motion (fBm) model over a wide range of length scales (Chakrabarti and Benguigui, 1998). We have shown that the profile of the observed kHz EMEs follows this model (Contoyiannis et al., 2005; Minadakis et al., 2012a).

The self-affine behavior can be quantitatively characterized by a single Hurst exponent, $H$, as the average heights difference $\langle y(x) - y(x+L) \rangle$ between two points on a profile increases as a function of their separation, $L$, like $L^H$ with $H \sim 0.75$, weakly dependent on the nature of the material, on the failure mode, and on the spatial scale of fracture. We have shown that the roughness of the profile of the observed kHz EM anomaly is in consistency with the aforementioned universal $H-$value (Kapiris et al., 2004; Contoyiannis et al., 2005; Minadakis et al., 2012a).

In the frame of the aspect of self-affine nature of faulting and fracture, the activation of a single fault should behaves as a "reduced image" of the regional seismicity, and a "magnified image" of the laboratory seismicity. We have shown that this happens: it has been shown that the populations of (i) EM-EQs included in an observed EM precursor associated with the activation of a single fault (SubSec. 4.1), (ii) EQs occurred on many faults included in a wide seismic region during a large time interval, and (iii) laboratory AE or EME pulses follow exactly the same relationship between frequency and event magnitude. This examination has been performed by means of traditional Gutenberg-Richter power-law, as well as in terms of a non-extensive function which is rooted in first principles of non-extensive statistical mechanics (Papadimitriou et al., 2008; Eftaxias, 2009; Minadakis et al., 2012a,b).

**4.3 Focus on extreme event features.**

A multidisciplinary analysis has also revealed that the kHz EM precursor possesses the following crucial characteristic of an extreme event (Karamanos et al., 2005, 2006; Kalimeri et a., 2008; Papadimitriou et al., 2008; Eftaxias et al., 2009; Eftaxias et al., 2010; Potirakis et al., 2012a, 2012b 2012c; Minadakis et al., 2012a; Eftaxias, 2012): (i) *High organization, high information content;* (ii) *Strong persistency*, indicating the presence of a positive feedback mechanism in the underlying fracto-EM mechanism that leads the systems out of equilibrium. (iii) *Existence of clear preferred direction of fracture activities*; (v) *Absence of any footprint of a second order transition in equilibrium or truncated- Lévy - walk type mechanism.*

In summary, it might be difficult for someone to accept that such emerged kHz EM anomalies are indeed seismogenic ones. However it is even more difficult to prove that they are not.

**5. Stage 3: Does the emergence of the kHz EM silence constitute a puzzling feature?**

As it was said in the previous section, recent high quality laboratory experiments strongly support the hypothesis that the observed EM silence before the EQ occurrence is also observed at the laboratory scale. However, accumulated evidence enhances the view that this silence is not a puzzling feature but the final precursory signal indicating the stage of the preparation of dynamical slip which results to the fast, even super-shear, mode; this is a slip mode which surpasses the shear wave speed.

**5.1 An approach by means of laboratory experiments and numerical studies.**

Recent laboratory experiments and numerical studies justify the observed pre-seismic EM silence by means of the sharp drop in the contact area at the peak stage, the behavior of the elastic moduli as damage increases, and the recently performed clarification that the recorded AE and EME, in general, are not two sides of the same coin. More precisely:

1. Park and Song (2013) have presented a new numerical method for the determination of contact areas of a rock joint under normal and shear loads. They report that, "at the peak stage, the normal dilation was initiated, which led to a sharp drop in the contact area. Approximately 53% of the surface area remained in contact, supporting the normal and shear loads. The active zone was partially detached, and the inactive zone was partially in contact. After the peak stage, the contact area ratio *decreased rapidly* with increasing shear displacement, and few inactive elements came into contact until the residual stage. At the residual stage, only small fractions, 0.3%, were involved in contact." *The rapid decrease of contact area ratio is in full agreement with the observed abrupt cease of the kHz EME. Thus, the last emerged kHz EM avalanche (electromagnetic earthquake) into the EM precursor may reveal the damage of the last strong asperity which sustains the system in the sticking regime.* The above mentioned scenario which bridges the observed kHz preseismic EME with recent experimental results and leads to the crucial hypothesis that the observed EM precursor mirrors the damage of a critical number of asperities that leads to the transition to the preparation of the final fast slip calls for further documentation. We recall that, McGarr and Fletcher (2003) and McGarr et al. (2010) suggest that stick-slip friction events observed in the laboratory and EQs in continental settings, even with large magnitudes, have similar rupture mechanisms.

2. As it was mentioned, stick-slip events rooted in the damage of a strong contacts are characterized by sudden shear stress drops that range from 10%-30% of the maximum frictional strength (Johnson et al., 2008). The occurred sharp drop in stress means that the strain resistance dramatically decreases. Elastic moduli, characteristics of the solid being considered, are the key parameters for defining relationships between stress and strain and evaluating strain resistance. Laboratory and theoretical studies show that the break of an element is associated by decrease of elastic modulus of damaged material (Amitrano and Helmstetter, 2006); the elastic modulus significantly decreases as damage increases, approaching to zero as the global fracture is approaching (Lin et al., 2004; Shen and Li, 2004; Chen, 2012). On the other hand, elastic moduli also constitute crucial parameters for the detection of AE or EME from material experiencing "damage". An increase of Young modulus and strength enhances the EME amplitude (Nitsan, 1977; Khatiashvili, 1984; Rabinovitch et al. 2002; Fukui et al., 2005). It might be concluded that the observed kHz EM gap just before the EQ occurrence is further supported by the above mentioned well established behavior of the elastic moduli.

3. The up to a few years ago view was that AE and the EME are two sides of the same coin. On the other hand, laboratory experiments in terms of AE were showing that this emission continues increasing up to the time of the final collapse. The combination of the above mentioned features supported the up-to-now impression that the appearance of EM silence just before the EQ occurrence is really a puzzling feature. However, recent accumulated experimental evidence enhanced some older ones indicating that the aforementioned well accepted view was false; *the recorded AE and EME, in general, are not two sides of the same coin.* Indeed, simultaneous laboratory measurements of AE and EME reveal the existence of two categories of AE signals (Yamada et al., 1989; Mori et al., 1994; Morgounov, 2001; Mori et al., 2004a, b, 2006, 2009; Mori and Obata, 2008; Lacidogna et al. 2010; Baddari and Frolov, 2010; Carpinteri et al., 2012):

(i) *AE signals which are associated with EME signals.* Both emissions are simultaneously generated during the creation of new fresh surfaces which is accompanied by the rupture of interatomic bonds and charge separation.

(ii) *AE signals which are not associated with EME signals. It has been proposed that this category of AE is rooted in frictional noises that appear during the rearrangements of the previously created fragments which are not accompanied by significant production of new surfaces.*

As it was mentioned, *laboratory studies reveal that strong AE and EME are simultaneously observed during stress drops occurred close to peak stage in the load versus time diagram* (Fukui et al., 2005; Lacidogna et al., 2011; Carpinteri et al., 2011; Carpinteri et al., 2012). *At this* stage *new fresh surfaces, which are accompanied by the rupture of interatomic bonds and charge separation*, are produced. Thus the simultaneous emission of strong kHz AE and EME is reasonable. However, during the last phase of the post-peak stage, that is the softening branch in the load versus time diagram, which is not accompanied by significant production of new surfaces, no EME is detected, while, on the contrary, the most intense AE emerges (e.g., Morgounov, 2001; Baddari and Frolov, 2010; Carpinteri et al., 2012).

*In summary, recent laboratory and numerical evidence exactly corresponds to the data obtained at the geophysical scale where EME is observed just before the EQ while there is no EME at the time of the EQ occurrence.*

**5.2 An approach by means of granular packings.**

We recall that there is evidence for the existence of a wide range of sliding velocities (or shear rates), even super-shear rupture modes that surpass the shear wave speed in the final stage of fast dynamical slip (Scholz, 2002; Xia et al., 2004, 2005; Coker et al., 2005; Ben-David et al., 2010). The appearance of fast sliding implies the existence of a kind of *"lubrication"* mechanism between fault plates.

Recent studies verify that gouge formation which behaves as bearings is found to be ubiquitous in brittle faults at all scales, and most slip along mature faults is observed to have been localized within gouge zones, while gouge included in various faults display similar characteristics (Chester and Chester, 1998; Sornette 1999, Wilson et al, 2005; Reches and Dewers, 2005; Baker and Warner, 2012). Wilson et al. (2005) propose that the observed fine-grain gouge is formed by dynamic rock pulverization during the propagation of a single EQ; a gouge zone is quickly developing with progressive slip reaching thicknesses larger than the height of the asperities, and further grain-size reduction occurs by systematic grain crushing due to amplified grain-contact stresses enhanced by the formation of stress-chains (Reches and Dewers, 2005 and references therein). In this case, the system can be regarded as granular matter that is sheared by the two surfaces (Kawamura et al., 2012 and references therein). Thus, tectonic faults are a characteristic example of shear failure in narrow zones (Åström et al., 2001; Alonso-Maroquin, et al., 2006). Numerical studies show that the so-called shear bands appear, for example, in granular packings (Åström et al., 2000; 2001) while there local "rotating bearings" are spontaneously formed (Åström et al., 2000). Many authors report the discovery of a self-similar space-filling bearing in which an arbitrary chosen sphere can rotate around any axis and all the other spheres rotate accordingly with negligible torsion friction (Baram et al., 2004; Verrato and Foffi, 2011; Åström and Timonen, 2012; Reis et al., 2012). We note that not only granular dynamics simulations but laboratory Couette experiments (Veje et al., 1999) demonstrate the spontaneous formation of bearings processes. Such a bearing-like mechanism has been proposed to explain the "lubrication" of the fault surfaces (e.g., Åström et al., 2000; Åström et al., 2001; Baram et al., 2004; Alonso-Marroquin et al, 2006). The above mentioned *"lubrication"* mechanism justifies the appearance of preseismic EM silence.

Now we focus on the duration of the observed EM silence. Two regimes for granular friction have been proposed *the quasi-static and dynamic regimes* (Midi, 2004; Da Cruz et al., 2005; Mizoguchi et al., 2006, 2009; Forterre and Pouliquen, 2008 ; Hayashi and Tsutsumi, 2010; Kawamura et al., 2012 and references therein). *Laboratory and numerical studies show that a time interval is needed for the formation of a shear band in the granular medium and thus for the transition from quasi-static to dynamic surface flow of a granular system.* Numerical studies reveal that this transition is characterized by intermittent local dynamic rearrangements and can be described by an order parameter defined by the density of critical contacts, namely, contacts where the friction is fully mobilized. Analysis of the spatial correlation of critical contacts shows the occurrence of "fluidized" clusters which exhibit a power-law divergence in size at the approach of stability limit, as predicted by recent models that describe the granular systems during static/dynamic transition as a multiphase system (Sharon et al., 2002 and references therein). Laboratory studies also show local rearrangements. For example, quantitative X-ray diffraction analyses indicate that strain localization and grain size reduction are also accompanied by changes in the nature and abundance of phases at rock localities (Boulton et al., 2012). Laboratory studies by means of acoustic measurements (Khidas and Jia, 2012) reveal that when a granular medium is sheared, the shear strain is essentially localized in a narrow zone location at the mid-height of the box where a shear band is formed. Such a shear localization zone exhibits distinct features compared to the rest of the medium, including extremely large voids and the presence of a highly anisotropic network of force gains (Khidas and Jia, 2012 and references therein). Welker and McNamara (2011) have studied a numerical simulation of granular assemblies subjected to a slow increasing deviator stress. They found that during the first half of the simulation, sliding contacts are uniformly distributed throughput the packing, but in the second half, they become concentrated in certain regions. This suggests that the loss of homogeneity occurs well before the appearance of shear bands.

*In summary, laboratory, theoretical and numerical studies indicate that the stage of preparation of the fast dynamical slip is associated with the appearance of a rolling-type "lubrication" mechanism of the included gouge between the fault surfaces. This phase is not accompanied by significant damage (breaking bonds) of brittle and strong material. A time interval is needed for the formation of a shear band in the granular medium and thus for the transition from quasi-static to dynamic surface flow of a granular system. The absence of kHz EM emission just before and at the time of the EQ occurrence is therefore fully justified.*

**5.3. The heat-flow paradox and the EM silence paradox: two sides of the same coin.**

One of the unresolved controversies in this field is a phenomenon that geophysicists call "*the heat-flow paradox*" (Sornette, 1999; Alonso-Marroquin, 2006 and references therein). To allow for large EQs, a fault should have a large friction coefficient so that it can restore a large amount of elastic energy and overpass large barriers. According to common sense, when two blocks grind against one another, there should be friction, and this should produce heat. Thus, large EQs should generate a large quantity of heat due to the rubbing of the two fault surfaces. However, measurements of heat flow during EQs were not able to detect the amount of heat predicted by simple frictional models. Calculations using the value of rock friction measured in the laboratory, i.e., a typical friction coefficient between 0.6 and 0.9, lead to overestimation of the heat flux. As an example, one refers in this context to the heat flow observations made around the San Andreas fault, which show that the effective friction coefficient must be around 0.2 or even less (Alonso-Marroquin, 2006 and references therein).

It might be concluded that the paradox of EM silence and the heat-flow paradox are two sides of the same coin. Both paradoxes originate from the appearance of low dynamical friction coefficient during the last stage of EQ generation, namely, the stage that prepares the final fast, even supershear, slip. We cannot deny the existence of preseismic EME because of the EM silence paradox, in the same sense that we cannot deny the existence of EQs because of the heat-flow paradox.

*It might be concluded that the systematically observed EM silence just before the EQ occurrence is a crucial feature of fracture process from the laboratory up to the geophysical scale constituting the last precursor of the imminent global instability and not a paradox feature.*

## 6. Why are the EM signals associated with small precursory strain changes but not with much larger co-seismic strains? -Shedding light from nanoscale plastic flow on the geophysical scale.

The general observation of strong intermittent precursory EM signals without co-seismic ones is actually important because strain changes are largest at the time of EQ. Any mechanism of EM precursors generation must explain why EM signals are associated with small precursory strain changes but not with much larger co-seismic strains. *We argue that a size-scale effect explains the aforementioned, considered as paradox, feature.* A size-scale effect is defined as a change in material properties which is rooted in a change in either the dimensions of an internal feature or structure or in the overall physical dimensions of a sample.

It is now well established that plastic flow is size depended, characteristically, flow stress or hardness are increasing with decreasing volume of material under load (Miguel et al., 2001; Dimiduk et al., 2006; Ward et al., 2009 and references therein). Plastic deformation in macroscopic samples is described as a *smooth* process occurring in an elastic continuum. However, recent experiments on micron-sized crystals reveal *step-like stress-strain* curves. Dislocation dynamic model suggests that the onset of plastic flow corresponds to a non-equilibrium phase transition, controlled by the external stress that separates a jammed phase, in which dislocations are immobile, from a flowing phase (Miquel et al., 2002). Plastic flow proceeds thorough a sequence of intermittent slip avalanches (Uchic et al., 2004; Richeton et al., 2005, 2006; Dimiduk et al., 2006; Miguel and Zapperi, 2006; Csikor et al., 2007; Dahmen et al., 2009; Zapperi, 2012). The resulted irreversible deformations intermittently change the microscopic material shape, while the isolated slip events lead to jumps in the stress-strain curves (strain bursts).

The statistics of the aforementioned discrete changes can reveal the underlying processes. Importantly, the emerging population of discrete slip events of microplasticity follows a scale-free (power-law) size distribution. On the contrary, in macroscopic sample plasticity appears as a smooth process. Therefore, a raised intriguing question refers to the nature of the cut-off which truncates scale-free behavior at large avalanches. More precisely one wonders (Csikor et al., 2007): If there is no intrinsic limit to the magnitude of dislocation avalanches, why do we not see them in deformation curves of macroscopic samples? Are the properties of dislocation avalanches truly universal?

Through ultra-precise nanoscale measurements on pure metal crystals loaded above the elastic-plastic transition, Dimiduk et al. (2006) directly determined the size of the emerged discrete slip events; the displacement events, $\Delta l$, follow a scale-free distribution with probability density function $p(\Delta l) \sim \Delta l^{-a}$ with $a \sim 1.5$. The founded scaling relationship is independent of sample size over the range examined as well as the gradually increasing stress over the range of the test, namely, there is no work-hardening effect for single slip-plane flow. Based on an alternative approach suggested by Newman (2005) the authors estimated a power-law slope of $1.60 \pm 0.02$ by a bootstrap method. On the other hand, a statistical characterization of intermittent plastic strain bursts has also been performed by means of AE. Dynamical processes associated with, nucleation, motion and emergence of dislocation groups and regular dislocations pileups (such as slip bands and cracks) on the crystal surfaces, cause AE. Experimental studies through AE have revealed that the plastic flow in crystalline solids is characterized by temporal intermittency. The emerging AE is consisted of a sequence of intermittent avalanches. The energy $E$ of the acoustic bursts follows a scale-free distribution, having a probability density function $p(E) \sim E^{-\kappa}$ with $\kappa \sim 1.5 - 1.6$ (Weiss and Grasso, 1997; Weiss and Marsan, 2003; Miguel et al., 2001; Zaiser and Moretti, 2005; Richeton et al., 2005; 2006). The scale-free behavior is extended up to over 8 decades. It is characteristic the absence of any cut-off. We pay attention to the finding that the exponents associated with the probability density function of both discrete slip events and AE events are practically identical. This implies that a fixed fraction of the work done by the external stresses during an elongation jump is released in the form of acoustic energy (Schwerdtfeger et al., 2007).

Dimiduk et al. (2006) conclude that the aforementioned results support an emerging view that *a statistical framework that creates a coarse-grained description of dislocation response is needed to bridge the gap between the behavior of individual dislocations and the ensemble of dislocations that govern macroscopic metal plasticity. Importantly,* Sethna et al. (2001) propose that the existence of a scale-free set of variables that describe deformation suggests that such a coarse-graining variable set really exists. *This assessment puts dislocation motion in the same class as EQs, sand pile avalanches, magnetic domain dynamics, and a wide variety of other dynamical systems. Dislocated nanocrystals are a model system for studying EQs generation; in analogy to plate tectonics, smooth macroscopic-scale crystalline glide arises from the spatial and time averages of disruptive EQ-like events at the nanometer scale* (Dimiduk et al., 2006).

Csikor et al. (2007) determine the distribution of strain changes during dislocation avalanches by combining three-dimensional simulations of the dynamics of interacting dislocations with statistical analysis of the corresponding behavior, and establish the dependence of this distribution on microcrystal size. More precisely, according to their study, the avalanche strain distributions obey the general form $P(s) = Cs^{-\tau} \exp\left[-(s/s_0)^2\right]$, where $C$ is a normalization constant, $\tau$ is a scaling exponent, and $s_0$ is the characteristic strain of the largest avalanches. The authors tested the robustness of the former equation in various physical situations and concluded that the distributions can be described with a universal exponent $\tau = 1.5$.

To elucidate the physical origin of the observed cut-off, Csikor et al. (2007) consider the proposition that during the progress of an avalanche, two processes reduce the effective stress upon the dislocations (Zaiser and Moretti, 2005; Zaiser, 2006): (i) Because of intrinsic hardening coefficient $\Theta$, a higher driving stress is needed to sustain the avalanche. The stress required to sustain-plastic flow increases with deformation, as if an additional back-stress $\sigma_b = -\Theta\gamma$ was building up inside the crystal. The back-stress opposes the propagation of large plastic avalanches, including a finite characteristic size (Miguel and Zapperi, 2006). (ii) In case of displacement-controlled deformation, the driving stress decreases due to relaxation of the elastic strain. Based on these considerations they conclude that $s_0 \propto bE/L(\Theta + \Gamma)$, where $\Gamma$ is the effective stiffness of the specimen-machine system (for a cubic compression specimen with rigid boundaries $\Gamma$ equals the elastic modulus $E$), $L$ is the characteristic specimen size, and $b$ the dislocation Bungers vector modulus. Rescaling the experimental data points by setting $s \rightarrow S = sL\Theta/bE$ and using a hardening coefficient $\Theta = E/1000$, Csikor et al. (2007) found that the scaled experimental data and simulated results are described by a single, universal scaling function $P(S) \sim S^{-3/2} \exp\left[-(S/0.6)^2\right]$. Therefore, their results demonstrate the *universality* of avalanche behavior in plastic flow and elucidate the cross-over between episodic and smooth plasticity.

The fact that the avalanche strains decease in inverse proportion to the sample size explains why it is difficult to observe strain bursts in macroscopic samples. On the contrary, in AE measurements, the acoustic energy is recorded. The energy release associated with a dislocation avalanche may be assumed to be proportional to the dissipated energy $e$, which is related to the strain by $e \approx \sigma sV$, where $\sigma$ is the stress and $V$ is the volume. Hence, the cutoff of the AE energy distribution is expected to increase with sample size as $e_0 \propto L^2$. On the other hand, a strong correlation between AE and EME events has been well documented demonstrating that *during the plastic flow (damage)* both AE and EME are radiated as two sides of the same coin (Hadjicontis et al., 2007 and references therein).

*The above mentioned considerations explain why it is easy to observe the precursory strong intermittent avalanche-like EME at the geophysical scale while it is not easy to observe the associated intermittent strain bursts.*

**7. Are the observed kHz EM signals in inconstancy with other precursors?**

At this point we refer to the argument that the absence of simultaneous geodetic or seismological precursors means that the observed kHz EM anomalies are not preseismic ones (Geller et al., 1997a). Recent research results show that this argument is not valid any more.

**7.1 Focus on the existence of simultaneous geodetic precursors.**

We recall that the observed kHz EM precursory activity is associated with the stick-slip mode (see Section 4). Slip fluctuates spatially because of pinning on local asperities (Perfettini et al., 2001). Sensitive and fast measurements have been performed by Nasuno et al. (1998) on sheared layers undergoing stick-slip motion with simultaneous optical imaging. Measurements of vertical displacements reveal dilation of material associated with each slip event. The hypothesis that these vertical displacements also cause deformations on the Earth's surface is reasonable. Consequently, the hypothesis that a precursory kHz EM activity should be consistent with other precursors rooted in the deformation of the Earth's surface is really reasonable. This requirement is fulfilled. Indeed, *Synthetic Aperture Radar interferometry (SAR)* is an imaging technique for measuring the topography of a surface, its changes over time, and other changes in the detailed characteristics of the surface. This method has demonstrated potential to monitor and measure surface deformations associated with EQs. Such deformations have also been reported in the case of the Athens EQ. Interferometric analysis of satellite ERS2 SAR images leads to the fault model of the Athens EQ (Kontoes et al., 2000). This model predicts the activation of two faults; the main fault segment is responsible for the 80% of the total seismic energy released, while the secondary fault segment for the remaining 20%. On the other hand, two strong avalanche-like kHz EM anomalies have been detected before the surface Athens surface EQ ($M_W = 5.9$, Sept. 7, 1999) with the following characteristics: The first and second anomaly lasted for 12 and 17 hours respectively with a cessation of 12 hours; the second anomaly ceased at about 9 hours before the EQ (cf. Fig. 6). Importantly, the observed kHz EM precursor is in agreement with the above mentioned findings performed by SAR or seismological techniques: the larger kHz EM anomaly, the second one, contains approximately 80% of the total EM energy released, with the first one containing the rest 20%. Notably, the same distribution holds for the entropy and information content of these two strong EM anomalies (cf. Fig. 6) (Eftaxias et al., 2001; Potirakis et al., 2012b).

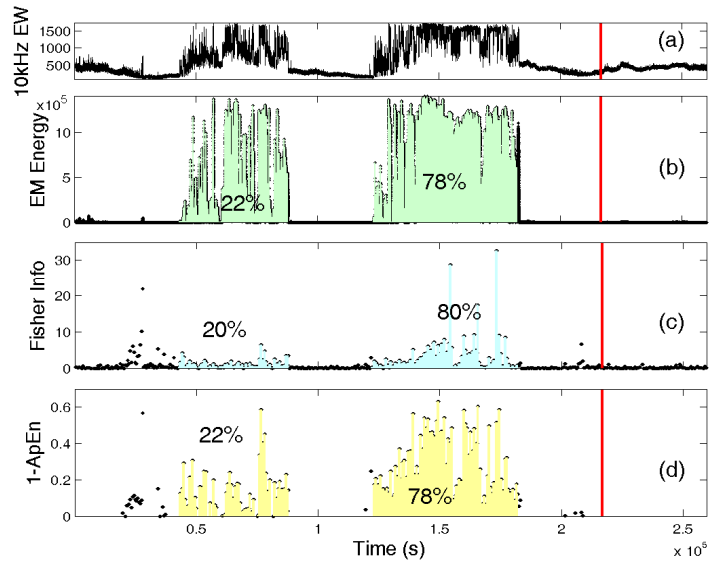

**Fig. 6** (a) The two strong impulsive bursts in the tail of the recorded preseismic kHz EM emission (10kHz, East-West, magnetic field strength in arbitrary units) prior to Athens EQ (please refer to Fig. 4). For the specific signal excerpt, the EM Energy (in arbitrary units) (b), the Fisher information (c) and approximate entropy (d) evolution with time are presented. The light green, blue and yellow colored areas indicate the energy, information and 1-ApEn corresponding to the two bursts, respectively. The first (left) burst is responsible for the 22% of the EM energy, the 20% of the Fisher information, and the 22% of the ApEn, while the second (right) for the 78%, 80%, and 78%, respectively. All graphs are time aligned, the vertical red line indicates the time of occurrence of the EQ.

**7.2 Focus on the existence of simultaneous seismological precursors.**

A fractured induced pre-seismic EM emission and the corresponding foreshock seismic activity also should be different manifestations of the same system. This requirement is valid, as well.

*Focus on the kHz EME.* A seismic data analysis in the case of the Athens EQ indicates that a two-event solution is more likely than a single event; two EQs emerged a few seconds one after the other (Eftaxias et al., 2001), while the moment ratio of the two seismic events is not in conflict with the aforementioned distribution of energy in the two emerged kHz bursts. Moreover, based on concepts of nonextensive statistical mechanics we have further elucidated the link between the precursory kHz EME the last stage of the impending EQ generation (Minadakis et al., 2012a,b). Especially, it has been shown that the statistics of regional seismicity behaves as a macroscopic reflection of the physical processes in the activated main fault, which are mirrored in the emerged kHz emission, as it would be expected by the fractal nature of fracture and faulting.

The above mentioned exceptional observational evidence, beyond any analysis, strongly: (i) enhances the hypothesis that the kHz EM precursor is associated with the fracture of asperities via a stick-slip mechanism, (ii) supports a strong relation between the observed kHz EM anomalies and the associated fault modeling in terms of interferometric and seismological analysis.

*Focus on the MHz EME.* We recall that it has been shown that this precursor can be described in analogy to the critical phase transitions in statistical physics (see Section 3). Based on the recently introduced method of natural time analysis we have shown that both the precursory MHz EM emission and the corresponding foreshock seismic activity present common signs of criticality supporting the hypothesis that the MHz EM emissions and the foreshock seismic activity are different manifestations of the same complex system at critical state (Potirakis et al., 2013).

*In summary, recent experimental evidence seems to shake down the negative argument under study concerning the seismic nature of the observed kHz EM anomalies.*

**8. On the systematically observed EM silence during the aftershock period.**

EME, as a phenomenon rooted in the damage process, should be an indicator of memory effects, as well. Indeed, laboratory studies verify that: during cyclic loading, the level of EME increases significantly when the stress exceeds the maximum previously reached stress level (Kaizer effect) (Lavrov, 2005 and references therein; Hadjicontis et al., 2005; Mori and Obata, 2008; Mavromatou et al., 2008). The existence of Kaizer effect can justify the aforementioned silence. The stress during the aftershocks period does not exceed the previously reached maximum stress level associated with the main shock occurrence.

**9. On the traceability of the EM precursors.**

A critical view often raised concerns the traceability of the fracture-induced EM emissions at the geophysical scale is that: "even if one accepts the generation of the EME before, and not at the EQ occurrence, an EM emission produced is the Earth's crust should be strongly attenuated by the Earth or, much more, by the sea before reaching the surface and being launched to the atmosphere." (Johnston, 1997).

First of all, we clarify that the observed EM precursors are associated with surface EQs that occurred on land or near coastline with magnitude ~6 or larger (e.g., Kapiris et al., 2002, 2003; Eftaxias et al., 2004; Karamanos et al., 2006). It is known that for an EQ with magnitude ~6 the fracture process extends to a radius of ~120km (Bowman et al., 1998). We argue that in this case the traceability of preseismic EME is justified. Indeed:

Accumulated evidence suggests that most of the released energy is consumed in creating the fault zone: (i) McGarr et al. (1979) conclude that "most of the released energy is consumed in creating the fault zone, with less than 1% being radiated seismically." (ii) Boler (1990) found that the energy of radiating elastic is smaller than 0.001 of the energy associated with *new areas.* (iii) Chester et al. (2005) conclude that energy required to create the *fracture surface area* in the fault is about 300 times greater than seismological estimates would predict for a single large EQ. *New surface areas* generated during an EQ is $S = 10^3 - 10^6 \, \text{m}^2$ for each $\text{m}^2$ of fault area. *We recall that during the formation of new surfaces EM radiations are emitted. Therefore, the hypothesis that a high amount of EME is radiated* *during the creation of the fault zone cannot be excluded.* On the other hand, a network of fracture traces having a fractal distribution in space is formed as the seismic event approaches. Fractals are highly convoluted, irregular shapes. The number of breaking bonds is dramatically higher in fractal fracture process in comparison to those of Euclidean fracture process. This situation justifies why a high amount of energy is consumed when the fault zone is created. The creation of the aforementioned network of traces / new surfaces forms a fractal network of EM emitters, namely, a Fractal Geo-Antenna which radiate in a cooperative way at the last stages of EQ preparation (Eftaxias et al., 2004). The observed precursory MHz – kHz EME are compatible to the notion of Fractal Geo-Antenna in the frame of the recently introduced Fractal Electrodynamics (Jaggard, 1990; Jaggard and Frangos, 2000), which combines fractal geometry with Maxwell's equations (Eftaxias et al., 2004). Interestingly: (i) The fractal dimension of the observed kHz EM precursors is $D = 1.2$, while a surface trace of a single major fault might be characterized by fractal dimension $D = 1.2$ (Sahimi et al., 1993; Sornette, 1991). (ii) *Optimal paths* play a fundamental role in fracture. Recently, Andrade et al. (2009) have explored the path that is activated once this optimal path fails and what happens when this new path also fails and so on, until the system is completely disconnected. The authors conclude that for all disorders the path along which all minimum energy paths fracture is a fractal of dimension $D = 1.22$.

The aforementioned concepts may also imply an answer why the nature plays meaningful $1/f$ music during the EQ preparation process. A huge amount of energy is consumed in risk-free (hazard-free) ruptures.

*It might be concluded that there is no reason why a high amount of fracture-induced EM emissions should not be directly launched through a Fractal Geo-Antenna to the atmosphere in the case of large surface EQs that occur on land.*

## 10. Discussion-Conclusions

This study is based on the consideration that a deeper understanding of the EM preseismic signals in terms recent results of basic science offers a way to achieve strict criteria for the characterization of the observed MHz and kHz EM anomalies as EQ precursors, and a better knowledge of the last stages of the EQ preparation process.

Based on a multidisciplinary analysis we conclude that the hypothesis that the following three stages model of EQ generation by means of preseismic fracture-induced MHz-kHz EM anomalies, *which is in full agreement with recent laboratory experiments,* cannot be excluded:

(i)	The initially launched preseismic MHz radiation, as it happens at the laboratory scale, is due to the fracture of the highly heterogeneous system that surrounds the formation of strong brittle and high-strength entities (asperities) distributed along the rough surfaces of the major fault. This appears as a complex cumulative process involving long-range correlations, interactions, and coalescence of cracking events. It is characterized by a *negative feedback mechanism* and *can be described in analogy with a thermal phase transition of second order.* The analysis by means of the method of critical fluctuations reveals the *"critical epoch"* during which the "short-range" correlations evolve into "long-range" ones, as well as the *epoch of localization of damage process along the major fault* in consistency with the crucial feature of "symmetry breaking" that characterizes a second order phase transition. A truncated Lévy walk type mechanism can organize the heterogeneous system to criticality. The criticality of the MHz EM emission has been verified by the recently introduced method of *natural time*. Importantly, it has been shown by means of natural time that the corresponding foreshock activity also shows criticality.

In summary, the aforementioned crucial features strictly define an observed MHz EME as a preseismic one. We suggest that the appearance of a well documented such anomaly *does not mean that the EQ is unavoidable.* Its launch implies that the "siege" of asperities has already been started. The abrupt emergence of strong avalanche-like kHz EM activity reveals the fracture of asperities, if and when the local stresses exceed their fracture stress (Contoyiannis et al., 2005). The absence of MHz EME during the main event is justified.

(ii)	The observed sequence of kHz EM avalanches which abruptly emerge in the tail of the preseismic EME, as it happens at the laboratory scale, originates in the stick-slip-like frictional stage preceding the preparation of final fast dynamic sliding. The sequence of electromagnetic bursts mirrors the sequential damage of asperities. This hypothesis is in consistency with: (i) recent laboratory studies;

(ii) universal structural patterns of fracture surfaces; (iii) the notion of self-affinity of fracture and faulting process; (iv) crucial characteristic of an extreme event; (v) with other precursors that are imposed by data from other disciplines, namely, seismology and SAR-interferometry.

In summary, the aforementioned crucial features strictly define an observed kHz EME as a preseismic one.

It might be difficult for someone to accept that a sequence of MHz and kHz EMEs which emerged one after the other within a short time interval, each of them fulfills the above mentioned strict criteria and are in consistency with other seismogenic precursors, is indeed a seismogenic one. However, it is even more difficult to prove that it is not. The hypothesis that the analysis of a sequence of MHz and kHz EM preseismic EME permits the step-by-step monitoring of the time evolution of the last stages of EQ generation cannot be excluded.

(iii)     The systematically observed EM silence just before the shock occurrence reveal the preparation of the final fast, even super-shear, sliding phase. We suggest that this, considered as enigmatic, feature is due to a *"lubrication"* mechanism which is organized by gouge included between fault surfaces similar to those appeared in *granular packings*. Numerical and laboratory studies support this hypothesis. We propose that the EM silence is another manifestation of the well known *heat-flow* paradox associated with the EQ occurrence. We cannot deny the existence of preseismic EME because of the EM silence paradox, in the same sense that we cannot deny the existence of EQs because of the heat-flow paradox. We emphasize that an EM silence appears just before the final rupture at recent laboratory experiments, as well.

In summary, the observed EM silence seems to constitute the last precursor of an impending EQ and not a puzzling feature.

The future will decide whether the presented three stage model is correct or not. However, the up to know laboratory experiments, numerical studies and theoretical consideration are in agreement with this model.

Attention has been devoted to other features considered as well as enigmatic ones: (i) The absence of simultaneous significant strain avalanches during the kHz EM avalanches observation is examined. We conclude that a *size-scale effect*, which has been resulted by recent high quality studies of plastic flow on micro-scale, fully explains the aforementioned, considered as inexplicable, feature of the study of EM presursors. (ii) The hypothesis that the systematically observed EME silence during the period of aftershocks occurrence EME is an indicator of *memory effects*, specifically the well known Kaiser effect, cannot be excluded. (iii) The distribution of the consumed energy in various stages of EQ preparation process, in connection with concepts sourced in the field of fractal electrodynamics seem to explain the, considered as problematic, traceability of the observed preseismic EME.

A very interesting research topic for the future would be *the comparative multidisciplinary* study of EME signals, both at MHz and kHz bands, recorded at different geological regimes of different seismological characteristics, and EQs characteristics, e.g., of different generation mechanisms. It could provide: (i) Useful information concerning future research in the field. (ii) The possibility for a *statistical verification* of the seismogenic nature of these signals. Indeed, a significant number of candidate precursor signals are necessary for a statistical evaluation. On the other hand, a reliable statistical identification of seismic precursors requires a two-step approach. First, an investigation "learning" step for the formation of a solid set of strict rules (required characteristics) for the reliable characterization of a candidate precursor as a valid one. Second, a recognition "test" step for the evaluation of future candidate precursors on an independent data set according to the previously established set of strict rules. We note that the above mentioned strict criteria for the characterization of an observed MHz or kHz EME as a precursor one offer for the first time such a set of rules. Therefore this work contributes to the first step of this approach. *We hope that this work will motivate research teams at different parts of the world to work on the acquisition and analysis of preseismic EME and eventually collect the required data for such a research in the future.* Then the second step will be feasible.

We should also note that other possible explanations of the involved natural processes preceding an EQ should also be investigated in the future. For example, the picture put forward by Papanikolaou et al. (2012) concerning the possible explanation of the "stick-slip" behavior could lead to interesting new results. According to that picture, "stick-slip" behavior can be also achieved (apart from increasing frictional forces) by increasing the rate of non-equilibrium relaxation channels that function to minimize the stress of the fault. In this context, strong relaxation mechanisms some days before the EQ events may also lead to stick-slip like behavior at the fault, but near the event these relaxation mechanisms cease to exist leading to a more random-like sliding (regular avalanche regime) and finally a very large EQ event.

Finally, the arguments presented in this paper for the justification of the observed EM silence just before and during the EQ occurrence, either referring to the recent laboratory experiments evidence, or through granular packings, imply that there is no significant production of new surfaces during the specific phase. The limited new surfaces that may be produced either are too limited to result to a sufficiently high level of EME so as to be detectable, or may even produce EME at higher frequency bands. Although there haven't been reported any EME at higher frequency bands (e.g., at GHz band) by the so far published laboratory experiments, it could be interesting to further investigate experimentally this possibility. Actually, our intention is to proceed to pilot installations of GHz detection systems in our field experimental infrastructure.

**References**

Alonso-Marroquin, F., Vardoulakis, I., Herrmann, H., Weatherley, D., and Mora, P.: Effect of rolling on dissipation in fault gouges, Phys. Rev. E, 74, 031306(1-10), 2006.
Amitrano, D., and Helmstetter, A.: Brittle creep, damage, and time to failure in rocks, J. Geophys. Res., 111, B11201(1-17), 2006.
Andrade, J. S., Oliveira, E. A., Moreira, A. A., and Herrmann, H. J.: Fracturing the optimal Paths, Phys. Rev. Lett., 103(22), 225503(1-4), doi:10.1103/Physrevlett.103.225503, 2009.
Åström, J., Herrmann, H., and Timonen, J.: Granular packing and fault zone, Phys. Rev. Lett., 84, 638-641, 2000.
Åström, J., Herrmann, H., and Timonen, J.: Fragmentation dynamics within shear bands-a model for aging tectonic faults, Eur. Phys. J. E., 4, 273-279, 2001.
Åström, J., and Timonen, J.: Spontaneous formation of dencely packed shear bands of rotating fragments, Eur. Phys. J. E., 35, 40(1-5), 2012.
Baddari, K., Sobolev, G. A., Frolov, A. D., and Ponomarev, A. V.: An integrated study of physical precursors of failure in relation to earthquake prediction, using large scale rock blocks, Ann. Geophys. 42(5), 771-787, doi: 10.4401/ag-3758, 1999.
Baddari, K., and Frolov, A.: Regularities in discrete hierarchy seismo-acoustic mode in a geophysical field, Annals of Geophys. 53, 31-42, 2010.
Baddari, K., Frolov, A., Tourtchine, V., and Rahmoune, F.: An integrated study of the dynamics of electromagnetic and acoustic regimes during failure of complex macrosystems using rock blocks, Rock Mech. Rock Eng. 44, 269-280, 2011.
Baker, K., and Warner. D.: Simulating dynamic fragmentation processes with particles and elements, Eng. Fracture Mech., 84, 96-110, 2012.
Baram, R, Herrmann, H., and Rivier, N.: Space-filling bearings in three dimensions, Phys. Rev. Lett., 92, 044301(1-4), 2004.
Baumberger, T., Caroli, C., and Ronsin, O.: Self-Healing Slip Pulses along a Gel/Glass Interface, Phys. Rev. Lett. 88, 075509(1-4), 2002.
Ben-David, O., Cohen, G., and Fineberg, J.: The dynamic of the onset of frictional slip, Science, 330, 211-214, 2010.
Boler, F.: Measurements of radiated elastic wave energy from dynamic tensile cracks, J. Geophys. Res., 95, 2593-2607, 1990.
Bouchaud, J-P., and Georges, A.: Anomalous diffusion in disordered media: statistical mechanisms, model and physical applications, Physics Reports, 195, 127-293, 1990.
Bouchon, M., Bouin, M-P., Karabulut, H., Toksoz, M., Dietrich. M., and Rosakis, A.: How fast is rupture during an earthquake? New insights from the 1999 Turkey earthquakes, Geoph. Res. Lett., 28, 2723-2726, 2001.
Bouchaud, E., and Soukiassian, P.: Fracture: from the atomic to the geophysical scale, J. Phys. D: Appl. Phys. 42, 210301, doi:10.1088/0022-3727/42/21/210301, 2009.
Boulton, C., Carpenter, B. M., Toy, V., and Marone, C.: Physical properties of surface outcrop cataclastic fault rocks, Alpine Fault, New Zealand, Geochem. Geophys. Geosyst., 13, 1-13, 2012.
Bowman, D., Quillon, G., Sammis, C., Sornette, A., and Sornette, D.: An observational test of the critical earthquake concept, J. Geophys. Res., 103, 24359-24372, 1998.
Carpinteri, A., Cornetti, P., and Sapora, A.: Brittle failures at rounded V-notches: a finite fracture mechanics approach, Int. J. Fracture, 172(1), 1-8, 2011.
Carpinteri, A., Lacidogna, G., Manuello, A., Niccolini, G., Schiavi, A., and Agosto, A.: Mechanical and Electromagnetic Emissions Related to Stress-Induced Cracks, SEM Exp. Techniq., 36, 53-64, 2012.
Chakrabarti, B., and Benguigui, L.: Statistical Physics of Fracture and Breakdown in Disordered Systems, Oxford University Press, Oxford, 1998.
Chang, J., Lockner, D., and Reches, Z.: Rapid acceleration leads to rapid weakening in earthquake-like laboratory experiments, Science, 338, 101-105, 2012.
Chauhan, V., and Misra, A.: Effects of strain rate and elevated temperature of electromagnetic radiation emission during plastic deformation and crack propagation in ASTM B 265 grade 2 Titanium sheets, J. Mat. Sci., 43, 5634-5643, 2008.
Chen, Y.Z.: A novel solution for effective elastic moduli of 2D cracked medium, Eng. Fracture Mechanics, 84, 123-131, 2012.
Chester, F., and J. Chester, J.: Ultracataclasite structure and friction processes of the Punchbowl fault, San Andreas system, California, Tectonophysics, 295, 199-221, 1998.
Chester, J., Chester, F., and Kronenberg, A.: Fracture surface energy of the Punchbowl fault, San Andreas system, Nature, 437, 133-136, 2005.
Coker, D., Lykotrafitis, G., Needleman, A., and Rosakis, A.: Frictional sliding modes along an interface between identical elastic plates subject to shear impact loading, J. Mech. and Phys. Solids 53, 884-922, 2005.
Contoyiannis, Y., and Diakonos, F.: Criticality and intermittency in the order parameter space, Phys. Lett. A., 268, 286-292, 2000.

Contoyiannis, Y., Diakonos, F., and Malakis, A.: Intermittent dynamics of critical fluctuations, Phys. Rev. Lett., 89, 035701(1-4), 2002.
Contoyiannis, Y., Kapiris, P., and Eftaxias, K.: A monitoring of a pre-seismic phase from its electromagnetic precursors, Phys. Rev. E, 71, 061123(1-14), 2005.
Contoyiannis, Y., and Eftaxias, K.: Tsallis and Levy statistics in the preparation of an earthquake, Nonlin. Proc. in Geophys., 15, 379-388, 2008.
Contoyiannis, Y.F., Nomicos, C., Kopanas, J., Antonopoulos, G., Contoyianni, L., and Eftaxias, K.: Critical features in electromagnetic anomalies detected prior to the L'Aquila earthquake. Physica A, 389, 499-508, 2010.
Contoyiannis, Y. F., Potirakis, S. M., and Eftaxias, K.: The Earth as a living planet: human-type diseases in the earthquake preparation process, Nat. Hazards Earth Syst. Sci., 13, 125-139, 2013.
Csikor, F., Motz, C., Weygand, D., Zaiser, M., and Zapperi S.: Dislocation avalanches, strain bursts, and the problem of plastic forming at the micrometer scale, Science, 318, 251-254, 2007.
Cyranoski, D.: A seismic shift in thinking, Nature, 431, 1032-1034, 2004.
Da Cruz, F., Eman, S., Prochnow, M., Roux, H-N., and Chevoir, F.: Rheophysics of dense granular materials: Discrete simulation of plane shear flows, Phys. Rev. E., 72, 021309(1-17), 2005.
Dahmen, K., Ben-Zion, Y., and Uhl, J.: Micromechanical model for deformation in solids with universal predictions for stress-strain curves and slip avalanches, Phys. Rev. Lett., 102, 175501(1-4), 2009.
Dimiduk, D., Woodward, C., LeSar, R., and Uchic, M.: Scale-free intermittent flow in crystal plasticity, Science, 312, 1188-1190, 2006.
Eftaxias, K., Kapiris, P., Polygiannakis, J., Bogris, N., Kopanas, J., Antonopoulos, G., Peratzakis, A., and Hadjicontis, V.: Signature of pending earthquake from electromagnetic anomalies, Geophys. Res. Lett., 28, 3321-3324, 2001.
Eftaxias, K., Kapiris, P., Dologlou, E., Kopanas, J., Bogris, N., Antonopoulos, G., Peratzakis, A., and Hadjicontis, V: EM anomalies before the Kozani earthquake: A study of their behavior through laboratory experiments, Geophys. Res. Lett., 29(8), 1228(1-4), doi:10.1029/2001GL013786, 2002.
Eftaxias, K., Frangos, P., Kapiris, P., Polygiannakis, J., Kopanas, J., Peratzakis, A., Skountzos, P., and Jaggard, D.: Review-Model of Pre-Seismic Electromagnetic Emissions in Terms of Fractal-Electrodynamics, Fractals, 12, 243-273, 2004.
Eftaxias, K., Panin, V.E, and Deryugin, Y. Y.: Evolution-EM signals before earthquakes in terms of meso-mechanics and complexity, Tectonophysics, 431, 273-300, 2007.
Eftaxias, K.: Footprints of nonextensive Tsallis statistics, selfaffinity and universality in the preparation of the L'Aquila earthquake hidden in a pre-seismic EM emission, Physica A, 389, 133-140, 2009.
Eftaxias, K., Athanasopoulou, L., Balasis, G., Kalimeri, M., Nikolopoulos, S., Contoyiannis, Y., Kopanas, J., Antonopoulos, G., and Nomicos, C.: Unfolding the procedure of characterizing recorded ultra low frequency, kHZ and MHz electromagetic anomalies prior to the L'Aquila earthquake as pre-seismic ones. - Part 1., Nat. Hazards Earth Syst. Sci.., 9, 1953-1971, 2009.
Eftaxias, K., Balasis, G., Contoyiannis, Y., Papadimitriou, C., Kalimeri, M., Athanasopoulou, L., Nikolopoulos, S., Kopanas, J., Antonopoulos, G., and Nomicos, C.: Unfolding the procedure of characterizing recorded ultra low frequency, kHZ and MHz electromagnetic anomalies prior to the L'Aquila earthquake as pre-seismic ones - Part 2, Nat. Hazards Earth Syst. Sci., 10, 275-294, 2010.
Eftaxias, K.: Are There Pre-Seismic Electromagnetic Precursors? A Multidisciplinary Approach, In Earthquake Research and Analysis - Statistical Studies, Observations and Planning 460 pages, InTech, March, doi: 10.5772/28069, 2012.
Forterre Y, and Pouliquen, O.: Flows of Dense Granular Media, Annual Review of Fluid Mechanics, 40, 1-24, 2008, doi: 10.1146/annurev.fluid.40.111406.102142.
Fukui, K., Ocubo, S., and Terashima, T.: Electromagnetic Radiation from rock during uniaxial compression testing: the effects of rock characteristics and test conditions, Rock Mech. Rock Eng., 38(5), 411-423, 2005.
Garcimartin, A., Guarino, A., Bellon, L., and Ciliberto, S.: Statistical properties of fracture precursors, Phys. Rev. Lett., 79, 3202-3205, 1997.
Geller, R., Jackson, D., Kagan, Y., and Mulargia, F.: Earthquakes cannot be predicted, Science, 275, 1616-1617, 1997a.
Geller, R., Jackson, D., Kagan, Y., and Mulargia, F.: Response in: Cannot earthquakes be predicted, Science, 278, 487-490, 1997b.
Geller, R.: Earthquake prediction: a critical review, Geophys. J. Int., 131, 425-450, 1997.
Girard, L., Amitrano, D., and Weiss, J.: Failure as a critical phenomenon in a progressive damage model, J. Stat. Mech., P01013(1-28), doi:10.1088/1742-5468/2010/01/P01013, 2010.
Girard, L., Weiss, J., and Amitrano, D.: Damage-cluster distributions and size effect on strength in compressive failure, Phys. Rev. Lett., 108, 225502(1-4), 2012.
Gokhberg, M., Morgunov, V., and Pokhotelov, O.: Earthquake Prediction, Seismo-Electromagnetic Phenomena, Gordon and Breach Publishers, Amsterdam, 193 pp., 1995.
Hadjicontis, V., Tombras, G. S., Ninos, D., and Mavromatou, C.: Memory effects in EM emission during uniaxial deformation of dielectric crystalline materials, IEEE Geosci. Remote Sens. Lett., 2(2), 118-120, 2005.
Hadjicontis, V., Mavromatou, C., Antsygina, T. N., Chishko, K. A.: Mechanism of electromagnetic emission in plastically deformed ionic crystal, Phys. Rev. B, 76, 024106(1-14), 2007.
Halasz, Z., Danku, Z., and Kun, F.: Competition of strength and stress disorder in creep rupture, Phys. Rev. E., 85, 016116(1-8), 2002.
Hayakawa, M., Atmospheric and Ionospheric Electromagnetic Phenomena Associated with Earthquakes, Terrapub, Tokyo, 996 pp, 1999.
Hayakawa, M., and Fujinawa, Y., Electromagnetic Phenomena Related to Earthquake Prediction, Terrapub, Tokyo, 667 pp., 1994.
Hayashi, N., and Tsutsumi, A.: Deformation textures and mechanical behavior of a hydrated amorphous silica formed along an experimentally produced fault in chert, Geophys. Res. Lett., 37, L12305(1-5), doi:10.1029/2010GL042943, 2010.
Helmstetter, A.: Is Earthquake triggering driven by small earthquakes?, Phys. Rev. Lett. 91, 058501(1-4), 2003.
Jaggard, D.: On fractal electrodynamics, in Recent Advances in Electromagnetic Theory, eds. H. Kritikos and D. Jaggard, Springer-Verlag, New York, 183-224, 1990.
Jaggard, D., and Frangos, P.: Surfaces and superlattices, in Frontiers in Electrodynamics, eds. D. Werner and R. Mittra, IEEE Press,1-47, 2000.
Johnston, M.: Review of electric and magnetic fields accompanying seismic and volcanic activity, Surveys in Geophysics, 18, 441-475, 1997.

Johnson, P., Savage, H., Knuth, M., Gomberg, J., and Marone, C.: Effects of acoustic waves on stick-slip in granular media and implications for earthquakes, Nature, 451, 57-60, 2008.

Kalimeri, M., Papadimitriou, K., Balasis, G., and Eftaxias, K.: Dynamical complexity detection in pre-seismic emissions using nonadditive Tsallis entropy, Physica A, 387, 1161-1172, 2008.

Kammer, D., Yastebov, V., Spijker, P., and Molinari, J-F.: On the propagation of slip at frictional interfaces, Trib. Lett., 48, 27-32, 2012.

Kapiris, P., Polygiannakis, J., Peratzakis, A., Nomikos, K., and Eftaxias, K.: VHF-electromagnetic evidence of the underlying pre-seismic critical stage, Earth Planets Space, 54, 1237-1246, 2002.

Kapiris, P., Eftaxias, K., Nomicos, K., Polygiannakis, J., Dologlou, E., Balasis, G., Bogris, N., Peratzakis, A., and Hadjicontis, V.: Evolving towards a critical point: A possible electromagnetic way in which the critical regime is reached as the rupture approaches, Nonlin. Proc.s in Geophys., 10, 511-524, 2003.

Kapiris, P., Eftaxias, K., and Chelidze, T.: Electromagnetic signature of prefracture criticality in heterogeneous media. Phys. Rev. Lett., 92(6), 065702(1-4), 2004.

Karamanos, K., Peratzakis, A., Kapiris, P., Nikolopoulos, S., Kopanas, J., and Eftaxias, K.: Extracting pre-seismic electromagnetic signatures in terms of symbolic dynamics, Nonlin. Proc. in Geophys., 12, 835-848, 2005.

Karamanos, K., Dakopoulos, D., Aloupis, K., Peratzakis, A., Athanasopoulou, L., Nikolopoulos, S., Kapiris, P., and Eftaxias, K., Pre-seismic electromagnetic signals in terms of complexity, Phys. Rev. E. 74, 016104(1-21), 2006.

Kawamura, H., Hatano, T., Kato, N., Biswas, A., and Chakrabarti, B.: Statistical physics of fracture, friction, and earthquakes, Rev. Modern Phys., 84, 839-884, 2012.

Khatiashvili, N.: The electromagnetic effect accompanying the fracturing of alcaline-halide crystals and rocks, Phys. Solid Earth, 20, 656-661, 1984.

Khidas, Y., and Jia, X.: Probing the shear-band formation in granular media with sound waves, Phys. Rev. E, 85, 051302(1-6), 2012.

Kontoes, C., Elias, P., Sykioti, O., Briole, P., Remy, D., Sachpazi, M., Veis, G. and Kotsis, I.: Displacement field and fault model for the September 7, 1999 Athens earthquake inferred from ERS2 satellite radar interferometry, Geophys. Res. Lett., 27(24), 3989-3992, 2000.

Kossobokov, V.: Testing earthquake prediction methods: the West Pacific short-term forecast of earthquakes with magnitude MwHRV5.8, Tectonophysics, 413, 2531, 2006.

Kumar, R., and Misra, A.: Some basic aspects of electromagnetic radiation emission during plastic deformation and crack propagation in Cu-Zn alloys, Materials Sci. Eng. A, 454-455, 203-210, 2007.

Kwapień, J., and Drożdża, S.: Physical approach to complex systems, Physics Reports, 515, 115-226, 2012.

Lacidogna, G., Manuello, A., Carpinteri, A., Niccolini, G., Agosto, A., and Durin, G.: Acoustic and electromagnetic emissions in rocks under compression, in: Proceeding of the SEM Annual Conference, Indianapolis, Indiana USA, 2010, Society for experimental Mechanics Inc, 2010.

Lacidogna, G., Carpinteri, A., Manuello, A., Durin, G., Schiavi, A., Niccolini, G., and Agosto, A.: Acoustic and electromagnetic emissions as precursors phenomena in failure processes, Strain, 47 (Suppl. 2), 144-152, 2011.

Lavrov, A.: Fracture-induced physical phenomena and memory effects in rocks: A review, Strain, 41, 135-149, 2005.

Lin, Q. X., Tham, L. G., Yeung, M. R., and Lee, P. K. K.: Failure of granite under constant loading, Int. J. Rock Mech. Min. Sci., 41, 362, 2004.

Lockner, D.: The role of acoustic emission in the study of rock fracture, Int. J. Rock Mech. Min. Sci. 30, 883-899, 1993.

Lockner, D.: Brittle fracture as an analog to earthquakes: Can acoustic emission be used to develop a viable prediction strategy? J. Acoustic Emission, 14, 88-101, 1996.

Lockner, D., and Madden, T.: A multiple-crack model of brittle fracture. Time-dependent simulations, J. Geophys. Res., 96(B12), 19643-19654, 1991.

Lockner, D., Byerlee, J., Kuksenko, V., Ponomarev, A., and Sidorin, A.: Quasi-static fault growth and shear fracture energy in granite, Nature, 350, 39-42, 1991.

Main, I., and Naylor, M.: Extreme events and predictability of catastrophic failure in composite materials and in the earth, Eur. Phys. J. Special Topics, 205, 183-197, doi:10.1140/epjst/e2012-01570-x, 2012.

Mantegna, R., Stanley, H.E.: Analytic approach to the problem of convergence of truncated Levy flights towards the Gaussian stochastic process, Phys. Rev. Lett., 73, 2946-2949, 1994.

Matsumoto, H., Ikeya, M., and Yamanaka, C.: Analysis of barberpole color and speckle noises recorded 6 and half hours before the Kobe earthquake, Jpn. J. App. Phys., 37, 1409–1411, 1998

Mavromatou, C., Tombras, G. S., Ninos, D., and Hadjicontis, V.: Electromagnetic emission memory phenomena related to LiF ionic crystal deformation, J. Appl. Phys., 103, 083518(1-4), doi:10.1063/1.2906346, 2008.

McGarr, A., Spottiswoode, S., Gay, N., and Ortlepp, W.: Observation relevant to seismic driving stress, stress drop, and efficiency, J. Geophys. Research, 84(B5), 2251-2261, doi:10.1029/JB084iB05p02251, 1979.

McGarr, A., and Fletcher, J.: Maximum slip in earthquake fault zones, apparent stress, and stick-slip friction, Bull. Seismol. Soc. Am., 93, 2355-2362, 2003.

McGarr, A., Fletcher, J., Boettcher, M, Beeler, N., and Boatwright, J.: Laboratory based maximum slip rate in earthquake rupture zones and radiated energy, Bull. Seismol. Soc. Am., 100, 3250-3260, doi:10.1785/0120100043, 2010.

Midi, G.: On dense granular flows, Eur. Phys. J. E, 14, 341-365, 2004.

Miguel, M-C., Vespignani, A., Zapperi, S., Weiss, J., and Grasso, J.-R.: Complexity in dislocation dynamics: model, Mater. Sci. Eng. A-Struct. Mater. Prop. Microstruct. Process., 309, 324-327. 2001.

Miguel, M.-C, and Zapperi, S.: Fluctuations in plasticity at the microscale, Science, 312, 1151-1152, 2006.

Miguel, M-C., Vespignani, A., Zaiser, M., and Zapperi, S.: Dislocation hamming and adrade creep, Phys. Rev. Lett., 89, 165501(1-4), 2002.

Minadakis, G., Potirakis, S. M., Nomicos, C., and Eftaxias, K.: Linking electromagnetic precursors with earthquake dynamics: an approach based on nonextensive fragment and self-affine asperity models, Physica A, 391, 2232-2244, 2012a.

Minadakis, G., Potirakis, S. M., Stonham, J., Nomicos, C., and Eftaxias, K.: The role of propagating stress waves in geophysical scale: Evidence in terms of nonextensivity, Physica A, 391(22), 5648-5657, doi:10.1016/j.physa.2012.04.030, 2012b.

Mizoguchi, K., Hirose, T., Shimamoto, T., and Fukuyama, E.: Moisture-related weakening and strengthening of a fault activated at seismic slip rates, Geophys. Res. Lett., 33, L16319(1-4), 2006.

Mizoguchi, K., Hirose, T., Shimamoto, T., and Fukuyama, E.: Fault heals rapidly after dynamic weakening, Bull. Seismol. Soc. Am., 99, 3470-3474, 2009.

Morgounov, V.: Relaxation creep model of impending earthquake, Anali Di Geophysica, 44, 369-381, 2001.

Mori, Y., Saruhashi, K., and Mogi, K.: Acoustic emission from rock specimen under cycling loading, Progress on Acoustic Emission VII, 173-178, 1994.
Mori, Y., Obata, Y., Pavelka, J., Sikula, J., and Lolajicek, T.: AE Kaiser effect and electromagnetic emission in the deformation of rock sample, DGZ-Proceedings BB 90-CD, EWGAE, Lecture 14, 157-165, 2004a.
Mori, Y., Obata, Y., Pavelka, J., Sikula, J., and Lolajicek, T.: AE Kaiser effect and electromagnetic emission in the deformation of rock sample, J. Acoustic Emission, 22, 91-101, 2004b.
Mori, Y., Sedlak, P.,and Sikula, J: Estimation of rock in-situ stress by acoustic and electromagnetic emission, in Advanced Materials Research, 13-14, 357-362, 2006.
Mori, Y., and Obata, Y.: Electromagnetic emission and AE Kaiser Effect for estimating rock in-situ stress, Report of the research Institute of Industrial Technology, Nihon University, No. 93, 2008.
Mori, Y., Obata, Y., and Sikula, J.: Acoustic and electromagnetic emission from crack created in rock sample under deformation, J. Acoustic Emission, 27, 157-166, 2009.
Nasuno, S., Kudrolli, A., Bak, A., and Gollub, J-P.: Time-resolved studies of stick-slip friction in sheared granular layers, Phys. Rev. E, 58, 2161-2171, 1998.
Newman, M.: Power laws, Pareto distributions and Zipf's law, Contemporary Physics, 46, 323-351, 2005.
Nitsan V.: Electromagnetic emission accompanying fracture of quartz-bearing rocks, Geophys. Res. Lett., 4(8), 333-336, 1977.
Papadimitriou, C., Kalimeri, M., and Eftaxias, K.: Nonextensivity and universality in the earthquake preparation process, Phys. Rev. E, 77, 036101(1-14), 2008.
Papanikolaou, S., Dimiduk, D., Choi, W., Sethna, J., Uchic, M., Woodward, C., and Zapperi, S.: Quasi-periodic events in crystal plasticity and the self-organized avalanche oscillator, Nature, 490, 517-521, 2012.
Park, S., Johnston, M., Madden, T., Morgan, F., and Morrison, H.: Electromagnetic precursors to earthquakes in the ULF band: A review of observations and mechanisms, Rev. Geophys., 31, 117-132, 1993.
Park J.-W., and Song, J.-J.: Numerical method for determination of contact areas of a rock joint under normal and shear loads, Int. J. Rock Mech. Min. Sci., 58, 8-22, 2013.
Peng, Z., and Gomberg, J.: An integrated perspective of the continuum between earthquakes and slow-slip phenomena, Nature Geoscience, 3, 599- 607, 2010.
Perfettini, H., Schmittbuhl, J., and Vilotte, J.: Slip correlations on a creeping fault, Geophys. Res. Lett., 28, 2133-2136, 2001.
Potirakis, G. Minadakis, and K. Eftaxias, Analysis of electromagnetic pre-seismic emissions using Fisher Information and Tsallis entropy, Physica A, 391, 300–306, 2012a.
Potirakis, S. M., Minadakis, G., and Eftaxias, K.: Relation between seismicity and pre-earthquake electromagnetic emissions in terms of energy, information and entropy content, Nat. Hazards Earth Syst. Sci., 12, 1179-1183, doi:10.5194/nhess-12-1179-2012, 2012b.
Potirakis, S. M., Minadakis, G., and Eftaxias, K.: Sudden drop of fractal dimension of electromagnetic emissions recorded prior to significant earthquake, Nat. Hazards, 64(1), 641-650, doi: 10.1007/s11069-012-0262-x, 2012c.
Potirakis, S. M., Karadimitrakis, A., and Eftaxias, K.: Natural time analysis of critical phenomena: the case of pre-fracture electromagnetic emissions, CHAOS, 23(2), xxxx-yyyy, 2013. [accepted for publication in Chaos: An Interdisciplinary Journal of Nonlinear Science, tentatively scheduled for publication in the June issue: vol. 23, no. 2]
Rabinowicz, E.: The nature of static and kinetic coefficients of friction, J. Appl. Phys., 27,1373-1379, 1951.
Rabinovitch, A., Bahat, D., and Frid, V.: Similarity and dissimilarity of electromagnetic radiation from carbonate rocks under compression, drilling and blasting, Int. J. Rock Mech. & Min. Sci., 39, 125-129, 2002.
Reches, Z.: Mechanisms of slip nucleation during earthquakes, Earth Planetary Sci. Lett., 170, 475-486, 1999.
Reches, Z., and Lockner, D.: Nucleation and growth of faults in brittle rocks, J. Geophys. Res., 99, 18159-18173, 1994.
Reches, Z., and Dewers, T.: Gouge formation by dynamic pulverization during earthquake rupture, Earth and Planetary Sci. Lett., 235, 361-374, 2005.
Reis, S., Araújo, N., Andrade, J., and Herrmann, H.: How dense can one pack spheres of arbitrary size distribution, Europhysics Letters, 97, 1804/1-5, 2012.
Richeton, T., Weiss, J., Louchet, F.: Breakdown of avalanche critical behaviour in polycrystalline plasticity, Nature Mater., 4, 465-469, 2005.
Richeton, T., Dobron, P., Chmelik, F., Weiss, J., Louchet, F.: On the critical behaviour of plasticity in metallic single crystals, Sci. Eng. A-Struct. Mater. Prop. Microstruct. Process., 424, 190-195, 2006.
Rubinstein, S.M., G. Cohen, and J. Fineberg: Detachment fronts and the onset of dynamic friction, Nature, 430, 1005-1009, 2004.
Rubinstein, S.M., G. Cohen, and J. Fineberg: Dynamics of precursors to frictional sliding, Phys. Rev. Lett. 98, 226103, 2007.
Rumi, De., and Ananthakrishna, G.: Power laws, precursors and predictability during failure, Europhys. Lett., 66, 715-721, 2004.
Sahimi, M., Robertson, M., and Sammis, C.: Fractal distribution of earthquakes hypocenters and its relation to fault patterns and percolation, Phys. Rev. Lett., 70, 2186-2189, 1993.
Sammis, C., and Sornette, D.: Positive feedback, memory and predictability of earthquakes, Proc. Natl. Acad. Sci. U.S.A., 99, 2501-2508, 2002.
Schiavi, A., Niccolini, G., Terrizzo, P., Carpinteri, A., Lacidogna, G., and Manuello, A.: Acoustic emissions at high and low frequencies during compression tests in brittle materials, Strain, 47, 105-110, 2011.
Scholz, C. H.: The mechanics of earthquakes and faulting, 2nd Ed. Cambridge University Press, Cambridge, 2002.
Schwerdtfeger, J., Nadgorny, E., Madani-Grasset, F., Koutsos., V., Blackford, J., and Zaiser, M.: Scale-free statistics of plasticity-induced surface steps on KCl single crystals, J. Stat. Mech., L04001(1-6), doi:10.1088/1742-5468/2007/04/L04001, 2007.
Sethna, J., Dahmen, K., and Myers, C.: Crackling noise, Nature, 410, 242-250, 2001.
Sharon, E., Cohen, G., and Fineberg, J.: Effects of crack front waves on dynamic fracture, Phys. Rev. Lett., 88, 085503(1-4), 2002.
Shen, L., and Li, J.: A numerical simulation for effective elastic moduli of plates with various distributions and sizes of cracks, Int. J. of Solids and Structures, 41, 7471-7492, 2004.
Shimamoto, T., and Togo, T.: Earthquakes in the Lab, Science, 338, 54-55, 2012.
Sornette, D.: Self-organized criticality in plate tectonics, in: Spontaneous Formation of Space-Time Structures and Criticality, edited by Riste, T. and Sherrington, D., 57–106, Kluwer Academic Publishers, 1991.
Sornette, D.: Earthquakes: from chemical alteration to mechanical rupture, Phys. Rep., 313, 237-291, 1999.
Sornette, D.: Critical Phenomena in Natural Sciences, Springer, 2000.
Stanley, H. E.: Introduction to Phase Transitions and Critical Phenomena, Oxford University Press, 1987.

Stanley, H. E.:Scaling, universality, and renormalization: Three pillars of modern critical phenomena, Reviews of Modern Physics, 71(2), S358-S366, 1999.

Tsutsumi, A., and Shirai, N.: Electromagnetic signals associated with stick-slip of quartz-free rocks, Tectonophysics, 450, 79-84, 2008.

Uchic, M., Dimiduk, D., Florando, J., and Nix, W.: Sample dimensions influence strength and crystal plasticity, Science, 305, 986-989, 2004.

Uyeda, S., Nagao, T., and Kamogawa, M.: Short-term earthquake prediction: Current status of seismo-electromagnetics, Tectonophysics, 470(3-4), 205-213, 2009

Veje, C., Howell, D. W., and Behringer, R. P.: Kinematics of a two-dimensional granular Couette experiment at the transition to shearing, Phys. Rev. E, 59, 739-745, 1999.

Verrato, F. and Foffi, G.: Apollonian packing as physical fractals, Molecular physics, 109, 2923-2928, 2011.

Wang, E.-Y., and Zhao, E.-L.: Numerical simulation of electromagnetic radiation caused by rock deformation and failure, Int. J. Rock Mech. & Min. Sci., 57, 57-63, 2013.

Ward, D., Farkas, D., Lian, H., Curtin, W., Wang, J., Kim, K.-S., and Qi, Y.: Engineering size-effects of plastic deformation in nanoscale asperities, Nature, 106, 9580-9585, 2009.

Weiss, J. and Grasso, J.-R.: Acoustic emission in single crystals of ice, J. Phys. Chem., B101, 6113-6117, 1997.

Weiss, J., and Marsan, D.: Three-dimensional mapping of dislocation avalanches: clustering and space/time coupling, Science, 299, 89-92, 2003.

Welker, P., and McNamara, A.: Precursors of failure and weakening in a biaxial test, Granular Matter., 13, 93-105, 2011.

Wilson, B., Dewers, T., Reches Z., and Brune, J.: Particle size and energetics of gouge from earthquake rupture zones, Nature, 434, 749-752, 2005.

Wyss, M.: Cannot earthquakes be predicted?, Science, 278, 487-490, 1997.

Xia, K., Rosakis, A., and Kanamori, H.: Laboratory earthquakes: the sub-Rayleigh-to-supershear rupture transition, Science, 303, 1859-1861, 2004.

Xia, K., Rosakis, A., Kanamori, H., and Rice, J.: Laboratory earthquakes along inhomogeneous faults; directionality and supershear, Science, 308, 681-684, 2005.

Yamada, I., Masuda, K., and Mizutani, H.: Electromagnetic and acoustic emission associated eith rock fracture, Phys. Earth Planet. Int., 57, 1570168, 1989.

Zaiser, M.: Scale invariance in plastic flow of crystalline solids, Advances in Physics, 54, 185-245, 2006.

Zaiser, M., and Moretti, P.: Fluctuation phenomena in crystal plasticity-a continuum model, J. of Stat. Mech., P08004, doi:10.1088/1742-5468/2005/08/P08004, 2005.

Zapperi, S.: Looking at how things slip, Science, 330, 184-185, 2010.

Zapperi, S.: Current challenges for statistical physics in friction and plasticity, Eur. Phys. J. B., 85, 329(1-12), 2012.